\documentclass[%
twocolumn,
superscriptaddress,
nofootinbib,
amsmath,amssymb,
aps,
prd,
]{revtex4-1}

\usepackage{graphicx}
\graphicspath{{.}{./data/files/}}

\usepackage{float}
\usepackage{bm}
\usepackage[hidelinks]{hyperref}
\usepackage{xcolor}

\usepackage[utf8]{inputenc}
\usepackage[T1]{fontenc}
\usepackage{slashed}

\usepackage{fullpage}
\usepackage{enumerate}

\usepackage{mathalfa,amssymb,amsthm}
\usepackage{amsmath}

\begin{document}


\title{Binary neutron star mergers with a subsolar mass star
}
\author{Maxence Corman}
\email{maxence.corman@aei.mpg.de}
\affiliation{Max Planck Institute for Gravitational Physics (Albert Einstein Institute), D-14476 Potsdam, Germany}
\author{William E. East}
\email{weast@perimeterinstitute.ca}
\affiliation{%
Perimeter Institute for Theoretical Physics, Waterloo, Ontario N2L 2Y5, Canada.
}%
\author{Jocelyn S. Read}
\email{jread@fullerton.edu}
\affiliation{%
Nicholas and Lee Begovich Center for Gravitational Wave Physics and Astronomy,
California State University Fullerton, Fullerton CA 92831 USA.
}%


\date{\today}

\begin{abstract}
While there are a number of proposed formation channels for subsolar mass
compact objects, including black holes formed primordially, or neutron stars
that form in collapsar disks, there have yet to be any conclusive observations
of such objects.  Motivated by the possibility that, if such objects exist,
gravitational waves from binary mergers may reveal them, we study binary
neutron star mergers where one star has a subsolar-mass in order to determine
how well such systems are described by current models, and when they could be
distinguished from a system with a subsolar-mass black hole.  We perform fully
general-relativistic simulations of a $1.7\ M_{\odot}$ star merging with a
$0.8\ M_{\odot}$ star, leading to tidal deformabilities of up to
$\mathcal{O}(10^4)$ for the latter, and quantify how this affects the merger
dynamics and associated gravitation and electromagnetic signals.  In this
regime, we find mass transfer between the stars, as well as significantly lower disruption frequencies.
Though this is not captured by current
gravitational waveform models, we conclude that this does not
significantly impact the sensitivity of current gravitational wave detectors to
these sources. Assuming design sensitivity of the LIGO and Virgo detectors,
we find no biases in the recovered intrinsic parameters for 
signal-to-noise ratios $\lesssim 100$.
We also find that the large deformabilities lead to a significant
increase in the amount of dynamically ejected matter compared to equal mass systems, exceeding
the predictions of current phenomenological models.
\end{abstract}

\maketitle

\section{Introduction}
\label{sec:introduction}
The first detection of a gravitational wave signal from a binary neutron star (BNS)
merger, GW170817~\cite{LIGOScientific:2017vwq}, was simultaneously observed as GRB170817A and subsequently observed across the
electromagnetic spectrum, inaugurating a new era in multi-messenger
astronomy~\cite{LIGOScientific:2017zic,LIGOScientific:2017ync} and exploration of the properties of matter
at high densities~\cite{LIGOScientific:2018cki,De:2018uhw,Radice:2018ozg,Raithel:2018ncd,
Capano:2019eae}.
Following this first detection, a second BNS gravitational wave event was observed, GW190425 \cite{LIGOScientific:2020aai}, unfortunately without an observed electromagnetic counterpart. In contrast to GW170817, the total mass of GW190425 was larger than
the mass of observed BNS systems in our galaxy,
illustrating the potential of gravitational wave observations to reveal new
populations of systems. Since 2019, despite the unprecedented sensitivity of the LIGO and
Virgo detectors, no clear BNS candidates have been reported. However, two recent sub-threshold BNS events, GW231109\_235456\cite{Niu:2025nha} and S250818k\cite{ligo:S250818k}, have been associated with potential electromagnetic counterparts~\cite{Li:2025rmj, Kasliwal:2025keb}. In addition, in November 2025, the LIGO-Virgo-KAGRA collaboration reported the candidate S251112cm\cite{ligo:S251112cm}, classified as a subsolar-mass compact binary merger. If astrophysical, the source frame chirp masses for both S250818k and S251112cm lie between $0.1$ and  $0.87 M_{\odot}$, which requires the lighter compact object to be below $1\ M_{\odot}$. The candidate S250818k additionally favors a second component classified as a $\gtrsim 1.0 M_{\odot}$ neutron star.

The observation of even a single subsolar mass neutron star would be an important 
discovery, since the supernova mechanism for forming neutron stars predicts typical masses ranging from 1.2 to 2 $M_{\odot}$
\cite{Kiziltan:2013oja,Lattimer:2000nx,Douchin:2001sv,Lattimer:2012nd,Suwa:2018uni}.
However, considerable uncertainty remains, especially at the mass range limits.
A recent claimed observation of a 0.77 $M_{\odot}$ neutron star within a supernova 
remnant \cite{Doroshenko:2022} suggests that subsolar mass neutron stars could be formed 
astrophysically. In addition to such tentative observational evidence, theoretical models 
also predict plausible formation channels for subsolar mass neutron stars, for example 
through fragmentation in gravitationally unstable collapsar disks~\cite{Metzger:2024ujc}, 
although significant uncertainties remain.

In this work, we consider a BNS system in which one of the component neutron stars 
has a mass 
smaller than $1\ M_{\odot}$. 
There have been numerous proposals for primordial formation scenarios
for subsolar mass black holes~\cite{Sasaki:2018dmp}, and one important question is how 
distinguishable a subsolar mass black hole would be from a neutron star 
in a binary merger.
Gravitational wave observations allow for the measurement of the
masses and tidal distortions of neutron stars through
their imprint on the gravitational waveform~\cite{Hinderer:2007mb,Flanagan:2007ix}, 
together providing direct constraints
on the equation of state (EOS). Smaller mass neutron stars tend to be more
easily deformed by tidal effects, as these objects are less
dense than their massive counterparts. EOS models predict that 
subsolar neutron stars can have dimensionless tidal deformabilities ranging from 
$\mathcal{O}(10^3)$ for a 
$1\ M_{\odot}$ up to  $\mathcal{O}(10^6)$ for a 0.1 $M_{\odot}$ star 
\cite{LIGOScientific:2019eut,Baldo:1997ag,Mueller:1996pm,Danielewicz:2008cm,
Agrawal:2005ix,Nazarewicz:1996}. 
The constraints on the tidal deformability of neutron stars from GW170817 
\cite{LIGOScientific:2018cki,De:2018uhw,Radice:2018ozg,Raithel:2018ncd,
Capano:2019eae} and those on neutron star radius, obtained from the NICER
X-ray telescope \cite{Miller:2019cac,Riley:2019yda,Miller:2021qha}, 
have ruled out a large fraction of the
EOSs. The observation of low-mass binary systems could provide some of the most
stringent constraints on the EOS, as the tidal effects in their
gravitational wave signals are more prominent.
Thus, detecting a low-mass neutron star would provide
tighter constraints on the EOS while offering insights
into alternative evolutionary pathways for forming ultra-light binaries, 
and challenging current assumptions
about neutron star populations.

Previous searches have attempted to look for subsolar mass systems 
while neglecting the tidal deformability of the system \cite{LVK:2022ydq,Nitz:2022ltl}.
Reference~\cite{Bandopadhyay:2022tbi} 
showed that
the loss in sensitive volume can be as high as 78.4\%
for an equal mass binary system of chirp mass 0.17 $M_{\odot}$, 
in a search conducted using binary black hole (BBH) template banks.
This loss was calculated under idealized conditions by comparing matches between 
templates and fiducial signals. In practice, gravitational wave signals are
buried in noise, making it likely that the actual sensitivity loss is 
higher. Therefore, if a nearby subsolar BNS system had merged, 
the signal might have been missed by previous searches due to reduced sensitivity 
from not accounting for tidal deformability. More recently, Ref.~\cite{Kacanja:2024hme} 
performed a novel search to explicitly look for subsolar BNS
systems to ensure none of the low-mass neutron stars
have escaped detection by previous search criteria. After 
performing a search using the public data from the third
observing run (O3) of advanced LIGO and Advanced
Virgo detectors, no new mergers were detected, and a new upper limit was placed 
on the merger rate of BNSs. 

However, both the abovementioned searches, and standard methods for extracting information about the 
properties of the binary system from gravitational wave detector data via a coherent Bayesian
analysis, rely on accurate gravitational wave approximants throughout the entire parameter space.
In principle, numerical-relativity (NR) simulations 
would be the method of choice for the description of a BNS waveform. 
However, such simulations come with high computational costs and thus 
typically only cover the late stage of the inspiral and merger 
\cite{Hotokezaka:2015xka,Hotokezaka:2016bzh,Lehner:2016lxy,
Kawaguchi:2018gvj,Kiuchi:2019kzt,
Foucart:2018lhe,Dietrich:2018phi,Ujevic:2022qle,Gonzalez:2022mgo,East:2019lbk}. As a result, over the years, the community has developed a range of waveform models
for BNS systems \cite{Abac:2023ujg,Haberland:2025luz,Dietrich:2017aum,Dietrich:2019kaq,Bernuzzi:2014owa,Akcay:2018yyh,Gamba:2023mww,Nagar:2018zoe,Hinderer:2016eia,Steinhoff:2016rfi}. One thing that they all have in common is that they are calibrated
to numerical simulations close to the merger, where
post-Newtonian (PN) theory breaks down. While different groups have produced
high-quality NR data for BNS configurations in which the individual stars are
spinning and covering mass ratios in the range $[1.0,2.0]$ (see, e.g., Table VII of
Ref.~\cite{Abac:2023ujg}), with several addressing the effect of high mass ratios on the 
dynamics and merger process \cite{Lehner:2016lxy,Dietrich:2016hky,Radice:2018pdn,
Bernuzzi:2020txg,Papenfort:2022ywx,Ujevic:2022qle}, 
none of these studies has, to our knowledge, produced
NR data for subsolar mass neutron star binaries with individual tidal
deformabilities greater than $\mathcal{O}(5000)$.
Therefore, estimating the sensitivity of searches to subsolar BNS system is subject to theoretical bias. Furthermore, a reliable analysis
of detected gravitational wave signals relies on an accurate theoretical 
description to cross-correlate measured gravitational strain
data with gravitational approximants throughout the entire
parameter space. If the waveform models used were inaccurate, 
this would lead to a systematic bias in the extraction of 
information from the observed data \cite{Dudi:2018jzn,Samajdar:2018dcx,Samajdar:2019ulq,
Read:2013zra}.
In fact, even for well-measured events such as GW170817, the inferred mass posteriors 
exhibit extended tails toward lower masses that fall outside the regime in which current 
models are calibrated (see Fig.~4 of Ref.~\cite{LIGOScientific:2017vwq}). The posterior 
sampling for GW231109\_235456\cite{Niu:2025nha} also extends into the subsolar mass regime. 
This further highlights the 
importance of developing waveform models that remain reliable across the full parameter 
space, including regions that have not yet been directly probed by observations.

In this paper, 
we follow the evolution of subsolar BNS systems with individual
tidal deformabilities of up to $\mathcal{O}(10^4)$ and try to answer the following
questions. i) What is the impact of having such large tidal deformabilities on
the merger dynamics and associated gravitational signal? We find that matter 
effects lead to noticeably lower disruption frequencies, in
the sensitive band of current gravitational wave detectors, which are are inaccurately described
by current gravitational wave models. ii) Do current gravitational wave models used in LIGO-Virgo-Kagra (LVK) analyses lead
to a significant loss of sensitivity? We find that the missing physics in current
gravitational wave models does not significantly impact the sensitivity of the current generation
of ground based detectors. For all the configurations considered, the mismatch between our
numerical simulations and state-of-the-art waveform models remains below $\sim 7 \times 
10^{-4}$. This is presumably because the signal detectable by current ground-based detectors contains
negligible power at merger frequencies. iii) Do current gravitational wave models lead to biases in
estimation of the source properties? We also perform parameter estimation and find no
biases in the effective tidal deformability of the binary. In particular,
assuming design sensitivity of
the LIGO and Virgo detectors, for network signal-to-noise ratios (SNRs) below 105, 
we find that 
all intrinsic parameters are recovered within $1\sigma$ (i.e. one standard deviation).

Although the mass alone of a subsolar compact object cannot unambiguously determine 
whether it is a neutron star or a black hole, the tidal deformability provides
a powerful discriminant. A confident measurement of a large tidal deformability would be 
incompatible with a black hole interpretation and thus would provide clear evidence for 
the presence of a neutron star. In this paper, we also explore to what extent such a 
distinction could be achieved with current gravitational wave detectors and find that assuming the smaller
object would be a black hole, the companion neutron star would have an inferred tidal
deformability in tension with constraints from other observations.

The remainder of the paper is organized as follows. Details about the numerical setup
and the different configurations considered are described in Sec.~\ref{sec:methods}.
We present a qualitative discussion of the merger dynamics in Sec.~\ref{sec:results},
and extract information about the ejecta and remnant properties in Sec.~\ref{sec:remnant}.
In Sec.~\ref{sec:hybridization}, we compare the extracted gravitational wave signals to a set of state-of-
the-art gravitational wave models and in Sec.~\ref{sec:PE}, we 
perform an injection study to understand possible
systematic biases. We conclude in Sec.~\ref{sec:conclusion}. We discuss the accuracy
of our simulations and provide more numerical details in Appendix~\ref{app:convergence}.
Throughout this paper we use physical units for ease of comparison, unless otherwise 
stated.

\section{Methods and configurations}\label{sec:methods}
\subsection{Numerical setup}
\label{sec:numerics}
We simulate BNS mergers by evolving the Einstein
equations coupled to hydrodynamics using the code described in Ref.~\cite{East:2011aa}. 
We discretize the Einstein field
equations in the generalized-harmonic formulation using
fourth-order accurate finite differences and time integration. We model the neutron star matter as a perfect fluid, and
evolve the general-relativistic Euler equations in conservative form using high-resolution shock-capturing techniques. 
We use the same methods and parameters for evolving BNS binaries as in
Ref.~\cite{East:2019lbk}.
Our simulations use box-in-box adaptive mesh refinement provided by
the PAMR library \cite{PAMR_online}. We typically use seven
levels of mesh refinement in our simulations, unless otherwise noted.
We provide details on numerical resolution and convergence in
Appendix~\ref{app:convergence}.

\subsection{Initial configurations}
\label{sec:initial_data}
We use quasi-circular BNS initial data constructed with
the Frankfurt University/Kadath (\texttt{FUKA}) Initial Data code suite
\cite{Papenfort:2021hod},
which is based on an extended version of the \texttt{KADATH} spectral solver library
\cite{Grandclement:2009ju}.
The set of hydrodynamical evolution equations
is closed by an EOS connecting pressure $p$
to specific internal energy $\epsilon$ and rest mass density $\rho$, i.e., 
$p=p(\rho,\epsilon)$. We focus on two different equations of state, namely cold
piecewise polytropic fits \cite{Suleiman:2022egw} approximating the SLy2 
\cite{Douchin:2001sv} and BSk21 \cite{Goriely:2010bm} EOSs.
The latter is a prototypical stiff EOS predicting a radius of $R_{1.4} \sim 12.59$ km
for a $1.4\ M_{\odot}$ neutron star, while the former is softer with 
$R_{1.4} \sim 11.76$ km \cite{Suleiman:2022egw}. 
These EOSs give maximum masses of $2.053 M_{\odot}$ for SLy2 and $2.274 M_{\odot}$ for BSk21
for non-spinning stars,
and are consistent with pulsar observations
\cite{NANOGrav:2019jur,Riley:2021pdl,van_Kerkwijk_2011,Fonseca:2016tux,Antoniadis:2013pzd},
and with both electromagnetic and gravitational wave
observations \cite{Margalit:2017dij,Radice:2017lry,LIGOScientific:2018hze,Pang:2021jta,
Coughlin:2018fis,Radice:2018ozg}
of the BNS event GW170817 \cite{TheLIGOScientific:2017qsa},
as well as the gravitational wave observations of GW190814 \cite{Tan:2020ics} and GW190425
\cite{Fasano:2020eum}.

Thermal effects are added to the zero-temperature
polytrope with an additional pressure contribution of the form
$p_{\rm th} = (\Gamma_{\rm th}-1)\rho\epsilon_{\rm th}$,
where $\epsilon_{\rm th}$ denotes the excess specific energy compared
to the cold value at the same density. We use $\Gamma_{\rm th} =1.75$,
motivated by studies comparing non-zero temperature EOSs such as Refs.~\cite{Yasin:2018ckc,Bauswein:2010dn}.

We focus on one non-spinning BNS configuration with a total mass of 
$M = 2.5\ M_{\odot}$ and a mass ratio $q=0.39$, corresponding to individual 
gravitational masses of $M_A = 1.8\ M_{\odot}$ and $M_B = 0.7\ M_{\odot}$. The resulting
chirp mass is $\mathcal{M} \sim 0.95\ M_{\odot}$, 
which lies on the edge of both the standard and subsolar mass compact
binary template banks used for searches in O4 \cite{LIGOScientific:2025yae, Hanna:2024tom}.
The value of the mass ratio was chosen to enhance tidal effects.
We choose 
an initial separation of $\sim 50$ km, which corresponds to an initial
gravitational wave frequency of 483 Hz, approximately 10 orbital periods before merger.
We perform eccentricity reduction as described in Ref.~\cite{Kyutoku:2014yba}, and reduce the 
initial orbital eccentricity to $\lesssim  10^{-3} $. 

For comparison, we also simulate an equal mass ($q=1$) BNS with the same total mass, 
using the BSk21 EOS. The initial separation is also 50 km, 
approximately 10 orbital periods before merger.

\begin{table*}[t]
\centering
\begin{tabular}{|c c c c c c c c c c c|}
 \hline
	EOS& $M_{\rm A}[M_{\odot}]$ & $M_{\rm B}[M_{\odot}]$ & $q$ &$R_{\rm A}[\rm{km}]$ & $R_{\rm B}[\rm{km}]$ & $\Lambda_A$ & $\Lambda_B$ & $\tilde{\Lambda}$ & $e$ & $f$ [Hz]\\
\hline
        BSk21& 1.8 & 0.7 & 0.39& 12.49 & 12.26 & 101 & 1.88 $\times 10^4$ & 1403 & $8 \times 10^{-4}$& 483\\
        SLy2 & 1.8 & 0.7 & 0.39& 11.34 & 12.03 & 42  & 1.50 $\times 10^4$ & 1066  & $10^{-3}$ & 483 \\
        BSk21 & 1.25 & 1.25 & 1.0& 12.53 & 12.53 & 1023  & 1023 & 1023  & $10^{-3}$ & 473 \\

\hline
\end{tabular}
\caption{Properties of our BNS simulations. The columns give the equation of state (EOS), 
the individual neutron star gravitational masses $M_{\rm A,B}$, the mass ratio $q$, 
the individual radii $R_{\rm A,B}$, tidal deformabilities $\Lambda_{\rm A,B}$, 
the effective tidal deformability of binary $\tilde{\Lambda}$, residual orbital eccentricity $e$, 
and initial gravitational wave frequency $f$. 
The neutron stars are non-spinning.  }
\label{tab:summary}
\end{table*}

\section{Results \label{sec:results}}

\begin{figure*}
        \includegraphics[width=0.99\textwidth,draft=false]{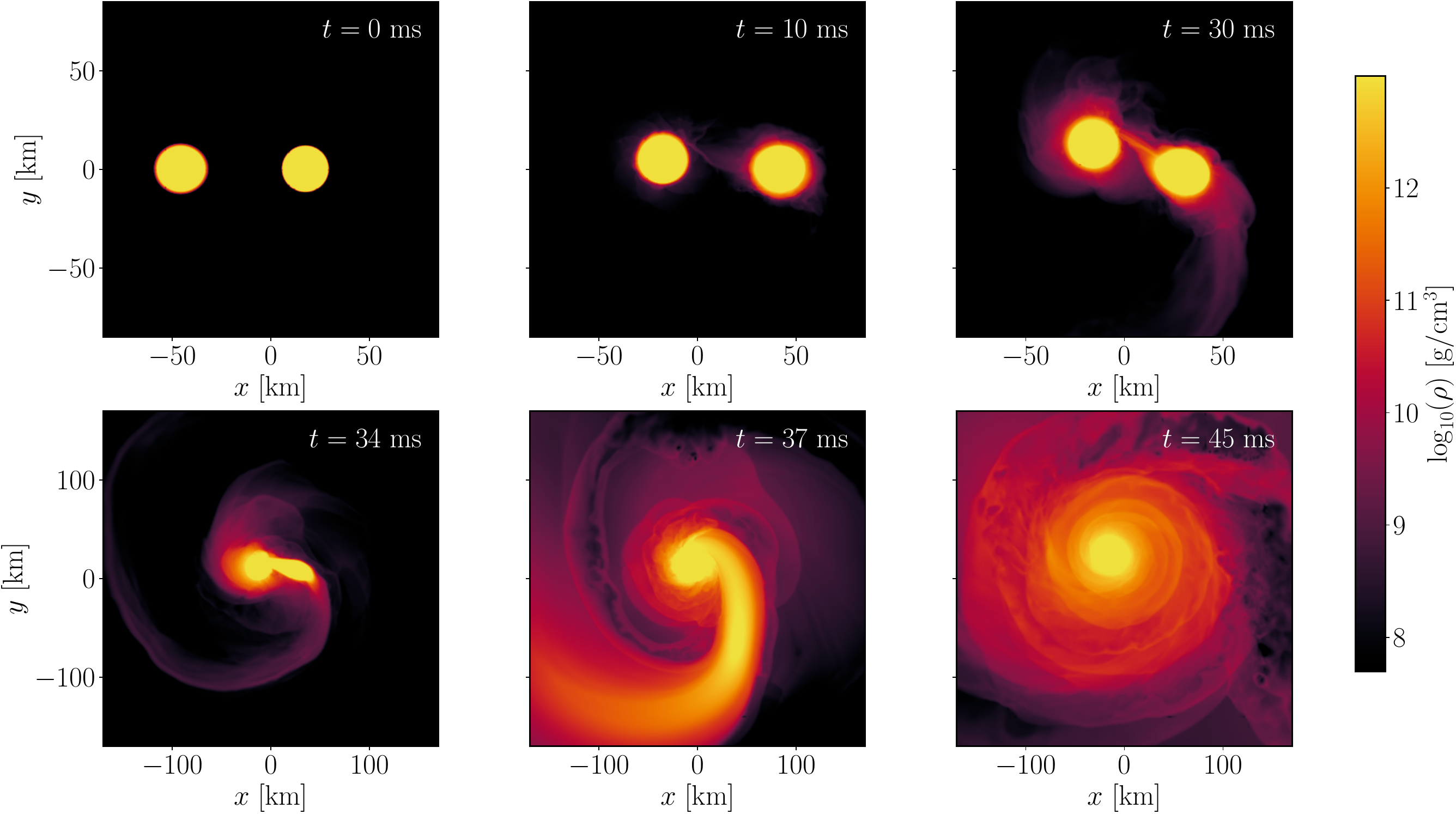}
        \caption{Rest-mass density profile in the equatorial plane on a logarithmic scale
	at different times for the BSk21 simulation.
        Note that the bottom row is zoomed out (by a factor of 2) compared to the top row.
\label{fig:density_snapshots}
}
\end{figure*}

\begin{figure}
        \includegraphics[width=0.49\textwidth,draft=false]{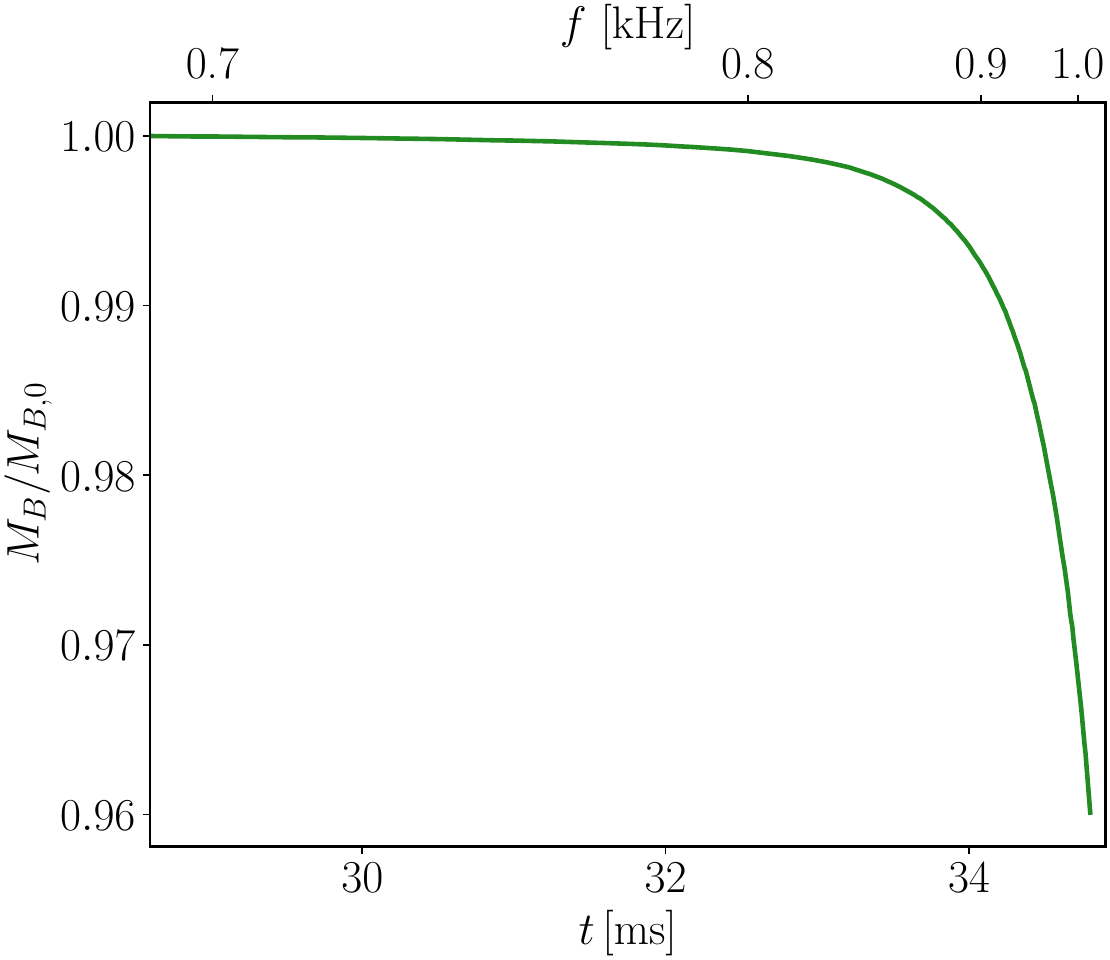}
	\caption{Change in the total rest mass surrounding the subsolar neutron star for the BSk21 simulation as a function of time. We also show the corresponding gravitational wave frequency at the top. We only show the mass evolution from $\sim 28$ ms onward as the mass does not vary much before then.  \label{fig:mass_transfer}
}
\end{figure}

We follow the evolution of the two types of BNS systems distinguished by their respective EOSs.
We summarize the properties of each binary in Table~\ref{tab:summary}.
In Fig.~\ref{fig:density_snapshots}, we present 2D snapshots of the density for the 
dynamical evolutions of the BSk21 system in the
EOS equatorial plane. In Fig.~\ref{fig:mass_transfer}, we show the rest-mass surrounding
the subsolar neutron star as a function of time and frequency.
Initially ($t=0$ ms or $f=483$ Hz), the objects are separated by a coordinate distance of $\sim 50$ km. The radii of the neutron stars are $R_{0.7} \sim 12.26$ km and $R_{1.8} \sim 12.49$ km, and their tidal deformabilities $\Lambda_{0.7} = 1.88 \times 10^4$ and 
$\Lambda_{1.8} = 101$. 
The second column (top row) of Fig.~\ref{fig:density_snapshots} shows the system at 
$t=10$ ms or $f=520$ Hz, i.e. $\sim 2.5$
orbits after the beginning of the simulation. At time $t \sim 30$ ms or $f\sim 720$ Hz, the smaller neutron star starts undergoing mass transfer. The
merger occurs at $t=35$ ms, or equivalently $\sim 990$ Hz, by which time the smaller neutron star
has lost roughly 4 $\%$ of its initial mass to the primary.
The final remnant is a long-lived massive neutron star that does
not collapse within the timescales of our simulations (up to $\sim 10$ ms post-merger).
The snapshot at $t=30$ ms shows that the smaller neutron star is highly deformed. 
This deformation, while
being coordinate dependent, is due to the large tidal deformability of the smaller neutron star,
which results in the star being disrupted at a lower frequency. The softer SLy2 EOS case 
undergoes a similar evolution, except the mass transfer and merger happen roughly $1.3$ ms later,
corresponding to a slightly higher merger frequency of $\sim 1050$ Hz.

In the following, we characterize the post-merger remnants, unbound material, and
the gravitational wave signals. Combining the gravitational waveforms from the
numerical simulations with analytical waveforms, we construct hybrid waveforms and
estimate the measurability of the dimensionless tidal deformability of the neutron stars
by advanced gravitational wave detectors.

\subsection{Ejecta and remnant properties}\label{sec:remnant}
In this section, we examine the post-merger distribution of matter.
To characterize the post-merger ejected matter, we use the integrated rest-mass density
$\rho$ residing outside some given radius
\begin{equation}
	M_0 (>r) = \int_{>r} \rho u^t \sqrt{-g} d^3 x
\end{equation}
where $u^t$ is the $t$ component of the fluid 4-velocity.
Post-merger, part of the rest mass will become unbound and
escape to infinity. We flag fluid elements as unbound if  $u_t < -1$ and the radial component of the
velocity is positive. From the value of $u_t$,
we can also determine the distribution of the rest mass $M_0$ over values of the
velocity at infinity $v_{\infty}$.

\begin{figure*}
        \includegraphics[width=0.49\textwidth,draft=false]{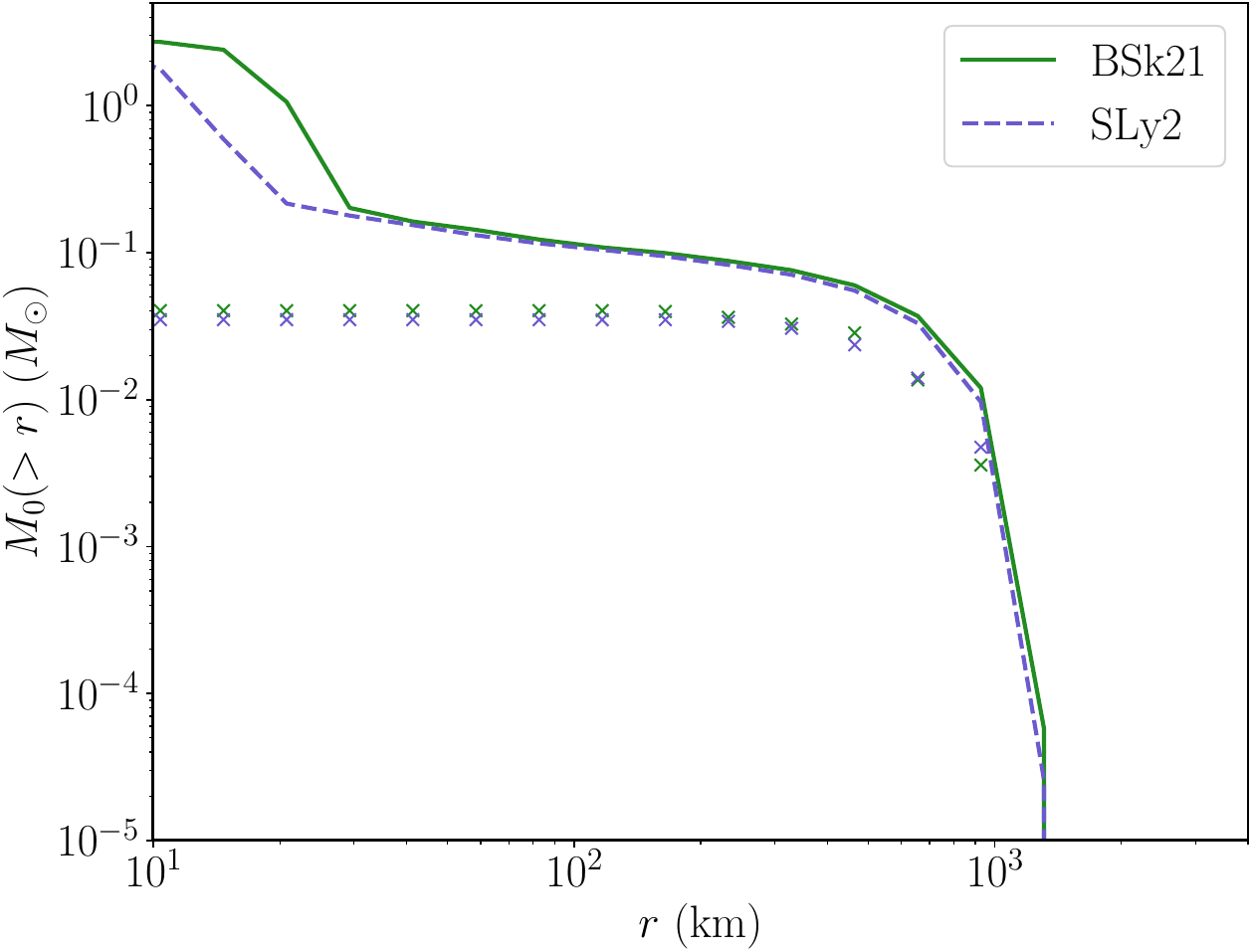}
        \includegraphics[width=0.49\textwidth,draft=false]{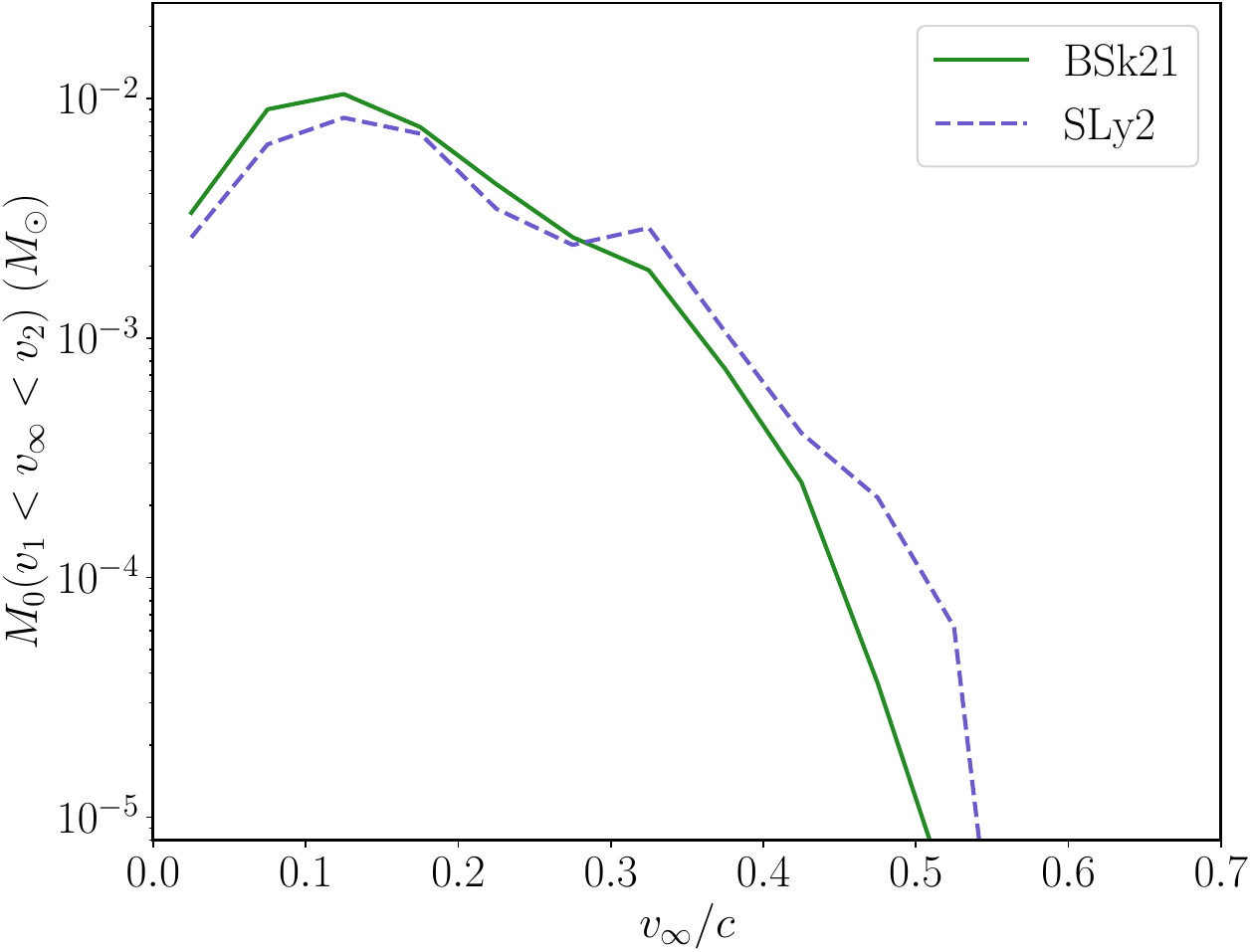}
        \caption{Left: The amount of rest mass outside a given coordinate radius from the center of mass for the post-merger remnant for mergers of BNSs with various EOSs. The solid curves show the total amount, while the points show just the
unbound rest mass. Right: The amount of unbound rest mass binned by the velocity at infinity, with each bin of width 0.05c.
\label{fig:matter}
}
\end{figure*}

Comparing the two different EOSs, we find in Fig.~\ref{fig:matter} that the stiffer BSk21 EOS
produces slightly more unbound matter than the softer SLy2 EOS: $4\times10^{-2}$ versus $3.5\times10^{-2}\ M_{\odot}$.
This trend in the ejecta differs from previous studies of BNS mergers 
\cite{East:2019lbk,Hotokezaka:2013,Radice:2016dwd,Baiotti:2016qnr}. 
One possible explanation is that, for equal-mass or mildly asymmetric binaries, 
shock heating is the dominant mass–ejection mechanism for neutron stars with soft EOSs,
whereas tidal interactions dominate for stiff EOSs.
However, the efficiency of shock heating decreases as the binary mass ratio departs from 
unity, while tidal effects become more important.
For stiff EOSs, where tidal interactions are already the primary driver of dynamical mass 
ejection, increasing mass asymmetry simply amplifies their role as also found in Ref.~\cite{Sekiguchi:2016bjd}.

We also investigate the dependence on mass ratio by performing an equal mass BSk21 
simulation with the same total mass as our unequal mass cases. In this setup, we find 
that the amount of unbound material is $\sim 1.3\times 10^{-3}\ M_{\odot}$, i.e, it decreases 
by a factor of approximately 30 compared to $q=0.39$. 
For comparison, we apply the fitting formulae from Ref.~\cite{Radice:2018pdn} and find 
good agreement in the equal-mass case. However, for the unequal-mass configuration, 
the fitting formula underestimates the amount of unbound material by a factor of about 6.

\subsection{Gravitational waves}
To study the gravitational wave signal, 
we decompose the polarizations $(h_{+},h_{\times})$ into $(\ell,m)$
multipoles as
\begin{equation}\label{eq:strain}
h_{+}(t) - i h_{\times}(t) = \frac{1}{D_L} \sum_{\ell= 2}^{\infty} \sum_{m=-\ell}^{\ell} h_{\ell m}(t) \, _{-2} Y_{\ell m} \left(\iota,\phi\right),
\end{equation}
where $D_L$ is the luminosity distance to source, $_{-2} Y_{\ell m}$ are the $s=-2$
spin-weighted spherical harmonics, and $\iota$ and $\phi$ are the inclination and
azimuthal angles that define the orientation of the binary with respect to the observer.
Each mode $h_{\ell m}(t)$ can be written in terms of a real time-domain amplitude 
$A_{\ell m}(t)$ and phase $\phi_{\ell m}(t)$ as 
\begin{equation}
	h_{\ell m}(t) = A_{\ell m}(t) e^{-i \phi_{\ell m}(t)},
\end{equation}
giving a gravitational wave frequency
\begin{equation}
	\omega_{\ell m}(t) = \frac{d}{dt}\phi_{\ell m}(t).
\end{equation}
The gravitational wave strain modes are related to the Newman-Penrose scalar $\Psi_4$ 
modes $\Psi_{4,\ell m}$ through $\ddot{h}_{\ell m}= \Psi_{4,\ell m}$. 
We extract the gravitational radiation by evaluating the Newman-Penrose scalar modes
$\Psi_{4,\ell m}$ at different extraction radii in the wave zone, and integrate 
the above equation to obtain the strain using the fixed frequency integration method \cite{Reisswig:2010di}.
The leading-order quadrupolar modes $\ell,m = 2, \pm 2$ are dominant in our case.

\begin{figure}
        \includegraphics[width=0.49\textwidth,draft=false]{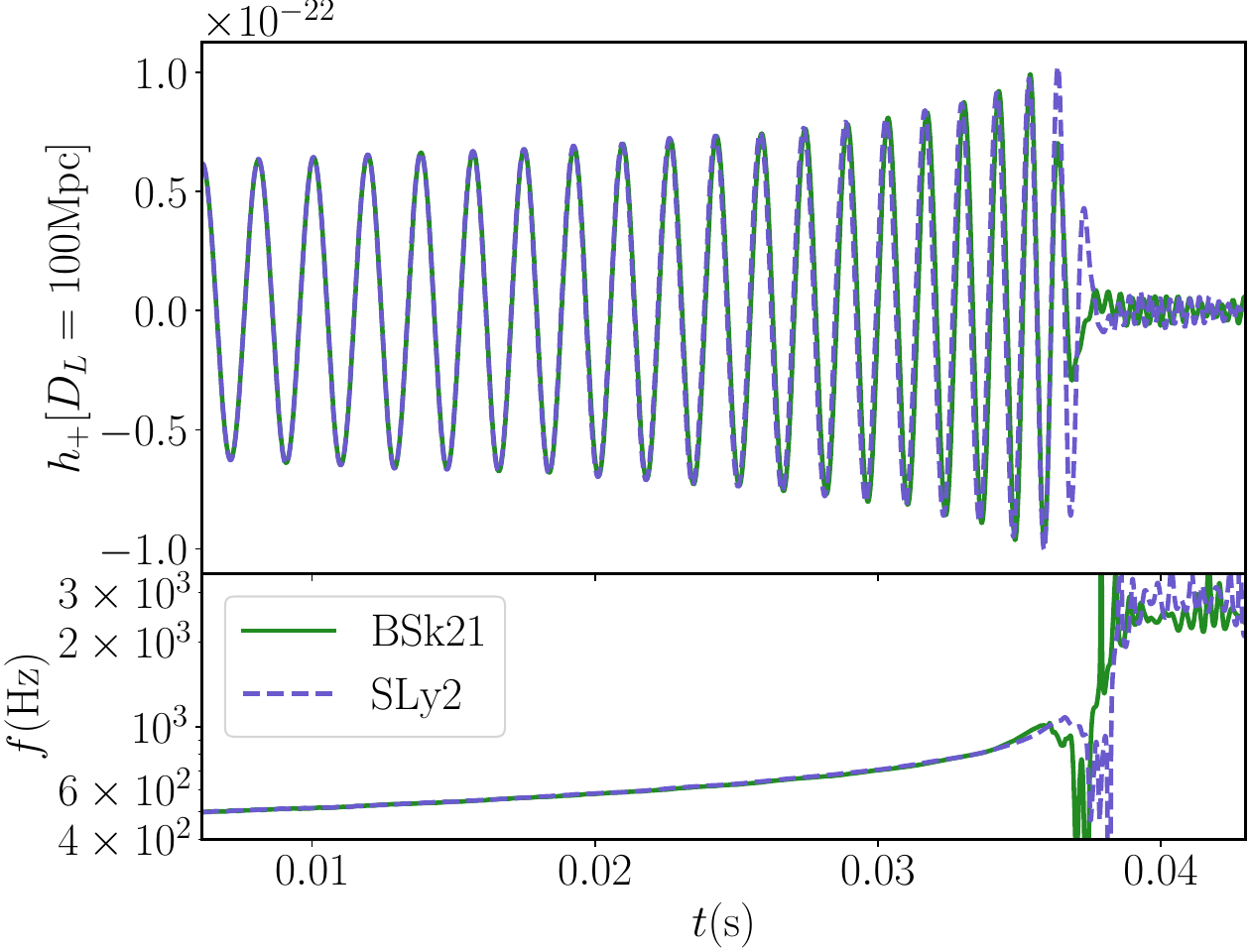}
	\caption{We show the plus polarization of the gravitational wave radiation (top) of the BNS mergers when observed face-on at 100 Mpc for two equations of state. The bottom panel shows the corresponding gravitational wave frequency. The waveforms are aligned in phase and time at a reference frequency of 500 Hz.
\label{fig:hplus}
}
\end{figure}

Figure~\ref{fig:hplus} 
shows the plus polarization $h_+$ for an optimally oriented, face-on source located
at $100$ Mpc for the two EOSs considered, aligned in phase and time at 500
Hz. We also show the gravitational wave frequency evolution. The peak frequencies, defined
as the instantaneous frequencies at which the amplitude of the strain is 
maximal, are 990 Hz
and 1050 Hz for BSk21 and SLy2, respectively.
For
comparison, the contact frequencies, defined as when the binary
separation equals the sum of the components’ radii, are $\sim 745$ and $\sim 810$ Hz, 
respectively (see appendix A of Ref.~\cite{Golomb:2024mmt} for a detailed definition).
Since the merger frequencies of these configurations lie within the sensitive band of ground-based detectors, we next study how well current gravitational wave models capture large tidal effects.

\subsection{Comparison with existing gravitational wave models}\label{sec:hybridization}
The numerical simulations produced in this paper consist of subsolar BNS 
simulations with large tidal effects and low eccentricity 
initial data (below $10^{-3}$). It is therefore natural to explore the performance of 
state-of-the-art inspiral-merger-ringdown BNS waveform models, which in contrast
to BBH models, need to include tidal effects. We focus on two major waveform-model families,
namely the SEOB and IMRPhenom models. The first waveform model we consider is
\texttt{IMRPhenomXAS\_NRTidalv3}, which employs
the frequency-domain aligned spin \texttt{IMRPhenomXAS}~\cite{Pratten:2020fqn} 
model as its BBH baseline, and 
augments it with tidal effects using the closed-form
\texttt{NRTidalv3}~\cite{Abac:2023ujg} prescription. The model is designed to
cover the ranges $m_{1,2} \in [0.5,3.0] \ M_{\odot}$ and 
$\Lambda_{1,2} \in [0,20000]$, a range well-suited to our study. 
However, we note that within the suite of numerical simulations that were used to 
calibrate the model, only two were unequal-mass systems
with a subsolar mass secondary (0.98 $M_{\odot}$ and 0.90 $M_{\odot}$, and with tidal
deformabilities $\sim 2600$ and $\sim 4600$, respectively). 
Within the SEOB family, we consider two models that incorporate
tidal effects in different ways. The first is
\texttt{SEOBNRv5\_ROM\_NRTidalv3}, which similarly to the \texttt{IMRPhenomXAS} model,
augments the BBH approximant \texttt{SEOBNRv5\_ROM}~\cite{Pompili:2023tna} 
with the \texttt{NRTidalv3} prescription. 
The second model, \texttt{SEOBNRv5THM}~\cite{Haberland:2025luz}, builds on the BBH 
approximant \texttt{SEOBNRv5HM}~\cite{Pompili:2023tna} and includes the same tidal 
information as 
\texttt{NRTidalv3}, apart from an extension of the dynamical tides formalism to 
spin-aligned neutron stars~\cite{Haberland:2025luz}. It also includes a calibrated merger 
time and a phenomenological model for the pre-merger waveform of the late inspiral which 
is calibrated to a large set of NR waveforms.
The \texttt{NRTidalv3} and \texttt{SEOBNRv5THM} not only differ in their treatment of the
merger, but also in the fact that in the former, tidal effects are added in the frequency
domain, while in the latter, they are incorporated in the time domain.
Given that noticeable differences between these models have been observed for mass ratios 
greater than 2, yielding mismatches up to $10^{-2}$~\cite{Haberland:2025luz},
it is particularly interesting to study which model best describes our numerical waveforms.

\begin{figure}
        \includegraphics[width=0.49\textwidth,draft=false]{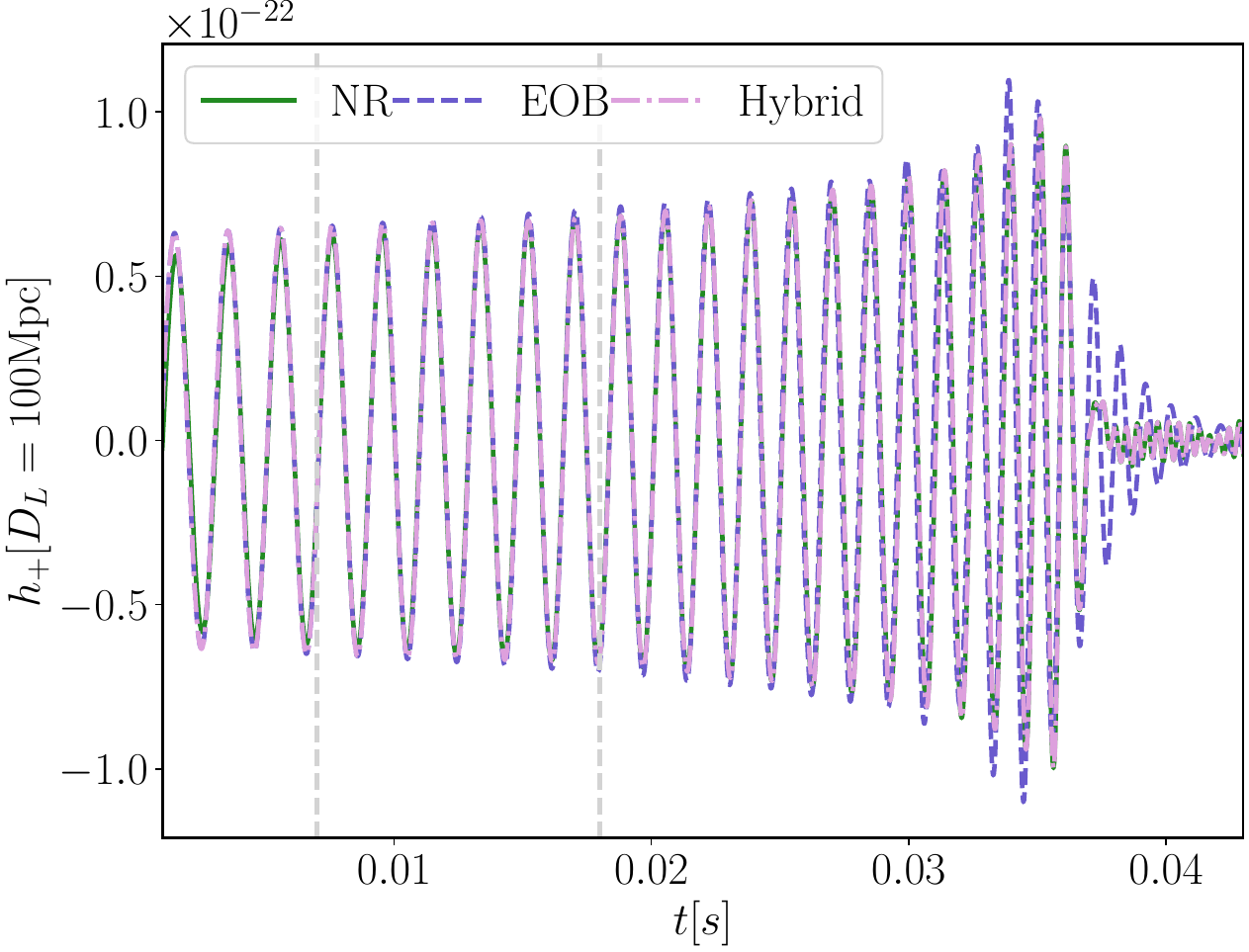}
\caption{Hybridization of the BSk21 configuration with
the \texttt{SEOBNRv5\_ROM\_NRTidalv3} model. The alignment interval is marked
by vertical dashed lines. The \texttt{SEOBNRv5\_ROM\_NRTidalv3} EOB waveform is
shown as a purple dashed curve and the NR waveform as a
green solid curve. The final hybrid combines the long inspiral 
from the EOB waveform, which includes several hundred
cycles (not shown in the figure), and the late inspiral, as well
as the post-merger phase of the NR waveform
\label{fig:hybrid}
}
\end{figure}

In order to make meaningful statements, the waveform models are compared against 
complete BNS waveforms that are constructed by stitching together our numerical waveforms,
covering the last $\sim 10$ orbits up to merger and extending through postmerger phase,
with inspiral waveforms (calculated from 20 Hz) obtained with respective 
waveform model. 
In particular, we hybridize the semi-analytical and NR waveforms by following the 
procedure outlined in Refs.~\cite{Dudi:2018jzn,Dietrich:2018uni,Hotokezaka:2016bzh}. 
We first align the semi-analytical (SA) and NR waveforms which employ the
same binary parameters by minimizing the integral
\begin{equation}
	\mathcal{I}(\delta t,\delta \phi) = \int_{t_i}^{t_f} dt \left| \phi_{\rm NR}
	- \phi_{\rm{SA}} (t + \delta t) + \delta \phi \right |
\end{equation}
where $\delta \phi$ and $\delta t$ are the relative phase and time shifts, and $\phi_{\rm NR}$
and $\phi_{\rm{SA}}$ denote the phases of the NR and in this case 
tidal EOB or IMRPhenom waveform. The alignment is done in a time window
$[t_i,t_f ]$ that corresponds to the frequency window $[500,565]$ Hz\footnote{The lower 
bound of the window is driven by the initial frequency of simulations and the upper bound is arbitrary, but we have verified that the results do not sensitively depend on the window we choose.}. Once the waveforms are aligned, we perform a smooth transition from the 
semi-analytical to the NR waveform within the interval:
\begin{equation}
	h_{\rm{hyb}}(t) =
    \begin{cases}
	    h_{\rm SA}(t^{\prime}) e^{i \phi} \hspace{3.55cm} t \leq t_i\\
	    h_{\rm NR}H(t) + h_{\rm SA}(t^{\prime}) e^{i \phi} \left[1-H(t)\right] \ t_i \leq t \leq t_f\\
	    h_{\rm NR}(t) \hspace{4.05cm} t\geq t_f
    \end{cases}       
\end{equation}
where $t^{\prime}= t + \delta t$, and $H(t)$ is the Hann window function,
\begin{equation}
H(t) := \frac{1}{2}\left[1 - \cos \left(\pi \frac{t-t_i}{t_f-t_i} \right) \right]
\ .
\end{equation}
In Fig.~\ref{fig:hybrid}, 
we present, as an example, the hybrid construction for the BSk21 configuration
and \texttt{SEOBNRv5\_ROM\_NRTidalv3},
with the alignment marked by vertical dashed lines.
The discrepancy between the numerical simulation and the waveform model 
in the merger and post-merger part of the waveform arises because \texttt{NRTidalv3} 
tapers the entire waveform past $7.5 f_{\rm est}$, where $f_{\rm est}$ is the
estimated merger frequency \cite{Abac:2023ujg}. 
This is also the reason for the sudden
decrease in amplitude of the waveform in the late inspiral \cite{Haberland:2025luz}.

\begin{figure*}
        \includegraphics[width=0.99\textwidth,draft=false]{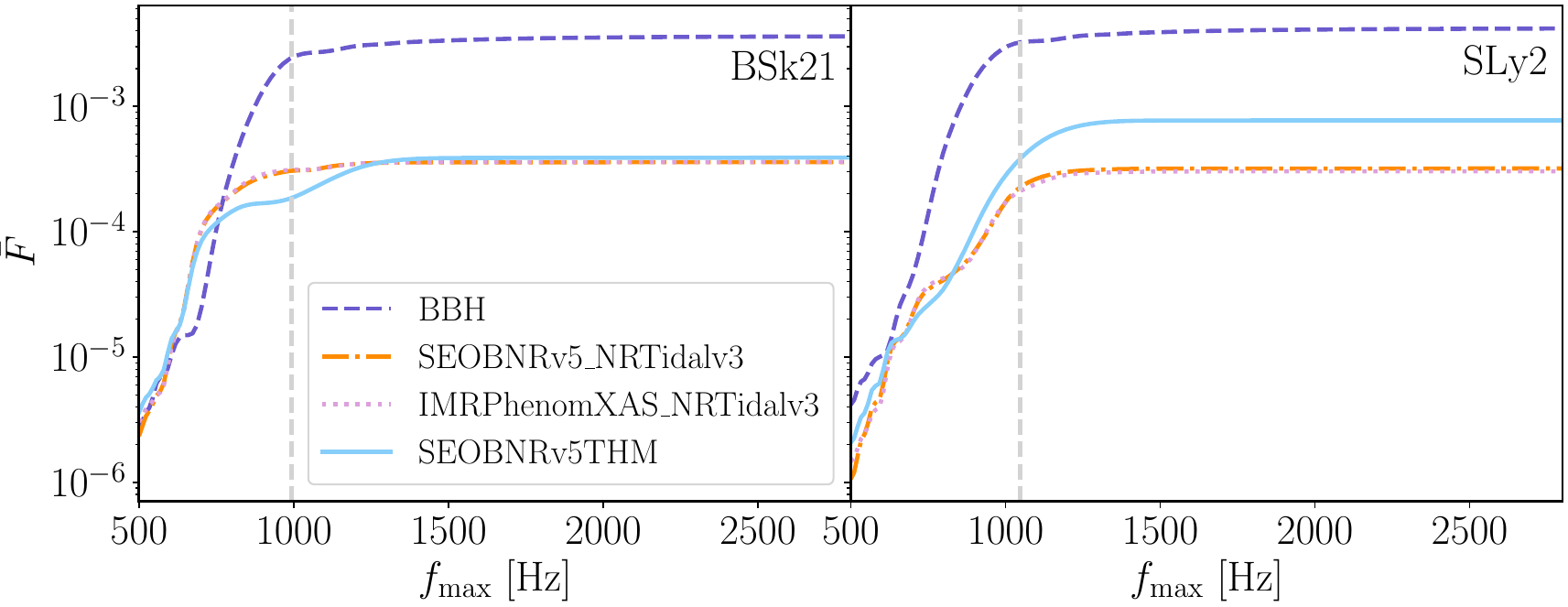}
        \caption{Mismatches between different waveform models and our hybrid waveforms 
	for the two different EOSs (BSk21 on the left, and SLy2 on the right), computed according to Eq.~\eqref{eq:mismatch} 
	with a fixed $f_{\rm min}= 20$ Hz and a variable maximum frequency 
	$f_{\rm max}$. 
	We mark the merger frequency with a vertical dashed line. 
\label{fig:mismatches}
}
\end{figure*}

Using these hybrid waveforms, we can now assess the gravitational wave models
by computing the mismatch between the hybrid waveforms and semi-analytical gravitational wave models.
We compute the mismatch according to
\begin{equation}\label{eq:mismatch}
	\bar{F} = 1 - \max_{\phi_c,t_c} \frac{\left(h_{{\rm SA}}(\phi_c,t_c)|h_{{\rm hyb}}\right)}{\sqrt{(h_{\rm SA}|h_{\rm SA})(h_{\rm hyb}|h_{\rm hyb}})}
 \ ,
\end{equation}
where $\phi_c$ and $t_c$ are arbitrary phase and time shifts, between approximants
and the hybrid waveforms. The noise-weighted overlap is defined as
\begin{equation}
	\left(h_1|h_2 \right) = 4 \mathcal{R} \int^{f_{\rm max}}_{f_{\rm min}} 
	\frac{\tilde{h}_1(f) \tilde{h}_2(f)}{S_n(f)} df
\ ,
\end{equation}
where the tilde denotes the Fourier transform, $S_n(f)$ is the spectral density of the detector
noise, and $f$ is the gravitational wave frequency (in the frequency domain). We use
the Advanced LIGO, zero-detuning, high-power (\texttt{aLIGO\_ZERO\_DET\_high\_P}) noise curve~\cite{aLIGO_noise} with a fixed $f_{\rm min}= 30$ Hz and a variable $f_{\rm max}$ ranging
from $500$ Hz up to 3000 Hz. We compute the mismatch for our hybrid waveforms against
the \texttt{IMRPhenomXAS\_NRTidalv3}, \texttt{SEOBNRv5\_ROM\_NRTidalv3}, 
\texttt{SEOBNRv5THM}, and \texttt{SEOBNRv5\_ROM} waveform models. 
Each hybrid is compared against the baseline
waveform model it was hybridized with, so that any contribution to the mismatch
arises solely from differences between the numerical simulations and the corresponding
waveform model. 
The results
for each EOS are shown in Fig.~\ref{fig:mismatches}. Generally, we find that
all waveform models, including the BBH waveform \texttt{SEOBNRv5\_ROM}, perform well. 
For all EOSs, the mismatch remains below $4 \times 10^{-3}$, even for maximum frequencies at or above the merger frequency. Note also that, the softer EOS gives mismatches
which are roughly of the same order of magnitude than the stiffer EOS, despite tidal 
effects being smaller. 
We emphasize that this is a non-trivial result since the parameters considered 
here lie outside the region of parameter space against which these waveform models were 
calibrated.
Another noteworthy result is that changing the BBH baseline between \texttt{SEOBNRv5\_ROM\_NRTidalv3} and \texttt{IMRPhenomXAS\_NRTidalv3} plays only a minor role in the mismatches, 
as was already pointed out in Ref.~\cite{Haberland:2025luz}. This is not
surprising as the two models agree in the low spin limit. We also note that
the \texttt{SEOBNRv5THM} mismatches are comparable to the \texttt{NRTidalv3} approximant.
The differences between the approximants are most likely due to the way the merger is 
handled in these two models, with \texttt{SEOBNRv5THM} 
merging for example later than \texttt{NRtidalv3} for the Sly2 system. Finally, for the 
BSK21  EOS we observe that the BBH approximant \texttt{SEOBNRv5\_ROM} yields smaller
mismatches than the tidal models up to roughly 750 Hz, i.e., before the onset of mass 
transfer. At first sight, this may appear counter-intuitive, since tidal effects are 
already present in the system. However, at these frequencies the tidal 
contributions to the phase evolution remain small, and differences between models are 
dominated by subtle amplitude and normalization effects rather than genuine tidal dynamics.
In particular, the BBH and tidal approximants are constructed and normalized differently. 
Since the mismatch is computed using normalized waveforms, small differences in amplitude 
modeling can influence the result.
As a consequence, the BBH model can yield slightly smaller mismatches in this regime 
despite omitting tidal physics. Once tidal effects become dynamically important, 
around the onset of mass transfer, the tidal models begin to outperform the BBH 
approximant, as expected. To better disentangle normalization effects from 
genuine waveform disagreement, we present two complementary diagnostics in Appendix~\ref{app:mismatches}.

Using these results, we can also provide a rough estimate of
the SNR above which
a particular waveform model will yield biases. In particular,
if two waveforms fulfill the criterion \cite{Flanagan:1997kp,Lindblom:2008cm,McWilliams:2010eq,Chatziioannou:2017tdw}
\begin{equation}
	\bar{F} < n_p/(2 \varrho^2)
\end{equation}
for a given power spectral density, where $n_p$ is total number of intrinsic parameters 
and $\varrho$ is the SNR, then they are deemed indistinguishable. Applying this to the BBH model, 
we find that the single detector SNR required to distinguish our
hybrid waveform from the BBH one is $\sim 41$ for the BSk21 and $\sim 39$ 
for the SLy2 configuration.
While this criterion is simple to evaluate, there are several problems that affect its 
usefulness in practice, and make it in general too conservative. Namely, if it is
violated, biases can, but need not arise \cite{Purrer:2019jcp} 
(see Ref.~\cite{Toubiana:2024car} for a suggestion on how to improve this criterion).
In the next section, we perform Bayesian inference, allowing us to properly answer beyond
which SNR a particular waveform model that is used as a template in parameter estimation 
yields biased posterior distributions for the hybrid signals.

\subsection{Parameter Estimation}\label{sec:PE}
We now perform an injection study to understand possible biases during parameter 
estimation for subsolar BNS systems. We consider a three-detector network
consisting of the LIGO and Virgo detectors.
Given $h_{\ell m}(t)$, the distance to the source, the angles $(\iota,\phi)$, $h_{+,\times}$ from Eq.~\eqref{eq:strain}, and the detector response functions $F_{+,\times}$, the
observed strain is:
\begin{equation}
	h(t) = h_+(t) F_+ (\alpha,\delta,\phi_p) + 
			h_{\times}(t) F_{\times} (\alpha,\delta,\phi_p)
\ ,
\end{equation}
where $F_{+,\times}$ depend on the source's sky location (right ascension $\alpha$ and
declination $\delta$) and the polarization angle $\phi_p$ of the source relative
to the interferometer.

We inject the BSk21 hybrid waveforms discussed in Sec.~\ref{sec:hybridization},
starting from 20 Hz, and do not consider the softer EOS of state further. 
We consider \texttt{SEOBNRv5\_ROM\_NRTidalv3} 
and the \texttt{SEOBNRv5\_ROM} BBH model. 
We do not consider the 
\texttt{IMRPhenomXAS\_NRTidalv3} model as the baseline BBH models agree in the low 
spin limit. Despite the promising mismatches, we also do not consider 
\texttt{SEOBNRv5THM} as it 
is not yet
implemented in the LALSuite package~\cite{lalsuite}.
We assume the design sensitivity of Advanced LIGO and Advanced Virgo. We
inject waveforms with a network SNR of $\varrho_{\rm net} \sim 35$, 
corresponding to 
an optimistic scenario similar to GW170817 \cite{TheLIGOScientific:2017qsa}.
In Table~\ref{tab:summary_extrinsic}, we list the extrinsic parameters of injected signals.
Because the high-frequency content of the post-merger portion of the gravitational wave reaches $\sim
2400$ Hz, we produce injections with a sampling rate of $16384$ Hz, which corresponds to a
high-frequency cutoff of $8192$ Hz. We recover with a sampling rate of $8192$ Hz and a
maximum frequency of $f_{\rm high}=3500$ Hz, i.e. above the waveform termination. 
All injections start at a frequency of 20 Hz, while
our Bayesian analysis uses a low cutoff frequency of $f_{\rm low} = 30$ Hz to generate
template waveforms. In all cases, we perform zero noise injections, as this allows us
to obtain posteriors that do not depend on a specific noise realization, therefore
isolating systematic errors.

We perform Bayesian inference
using \texttt{Bilby}~\cite{Ashton:2018jfp,Romero-Shaw:2020owr,Smith:2019ucc}
and analyze 256 s of data. 
We use a multi-banded likelihood calculation to speed up the computation and
the sampling algorithm used is nested sampling with the \texttt{Dynesty}
sampler \cite{Speagle:2020}. 
For the analysis, we assume uniform prior distributions in the interval
$[0.5, 1000]\ M_{\odot}$ for component masses and $[-0.05,0.05]$
for both dimensionless aligned spins. We also assume a uniform prior distribution for the
individual tidal deformabilities between $[0,2\times10^4]$. 
For all other parameters, we use priors similar
to other gravitational wave analyses; see, e.g., Ref.~\cite{LIGOScientific:2018hze}.
As a check 
that our pipeline works properly, we additionally inject a 
\texttt{SEOBNRv5\_ROM\_NRTidalv3} waveform constructed using the same parameters
as for the hybrid construction of BSk21 with that model, and we recover this injection
with the same model. We label this additional injection 
\texttt{SEOBNRv5\_ROM\_NRTidalv3}\_same\_inj and do not expect to see biases for this 
setup, which will provide a useful point of comparison.

\begin{figure*}
        \includegraphics[width=0.99\textwidth,draft=false]{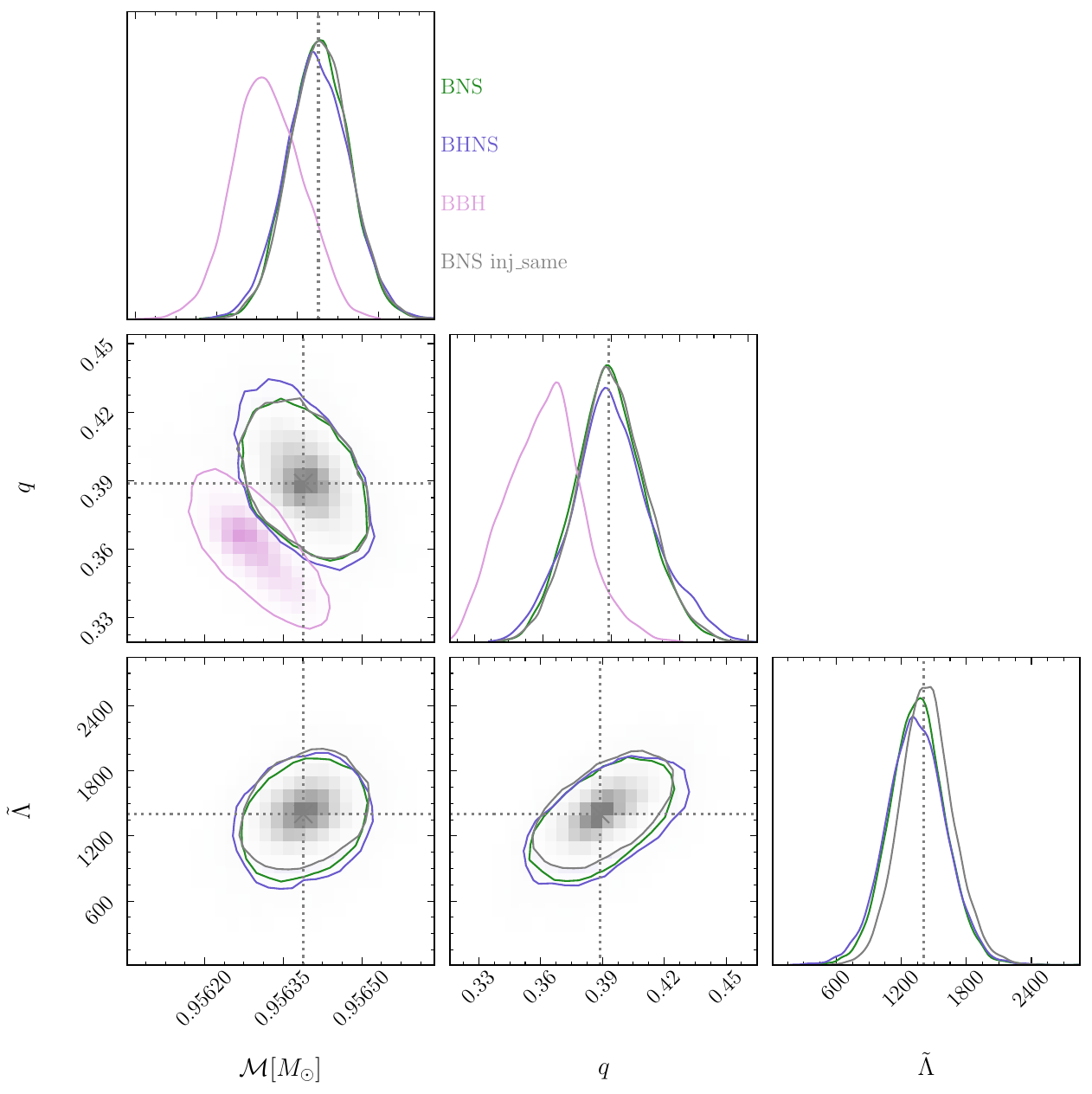}
\caption{The marginalized 1D and 2D posterior probability distributions for chirp mass 
$\mathcal{M}$, mass ratio $q$, and tidal deformability $\tilde{\Lambda}$ for the BSk21 
configuration. Recoveries labeled by BNS (green), 
BHNS (blue), and BBH (pink) use 
different approximations to recover our hybridized waveform, which is a combination of the 
\texttt{SEOBNRv5\_ROM\_NRTidalv3} model and our numerical simulation. The 
BNS inj\_same (gray) uses for injection and recovery 
the \texttt{SEOBNRv5\_ROM\_NRTidalv3} model with the same source parameters as the hybrid.
The latter serves as validation for our injection setup. The injected source parameters
are marked with dotted lines and crosses. Contours show the 90\% credible regions in 2D 
and the corresponding intervals in the 1D posteriors.
\label{fig:PE}
}
\end{figure*}

In Fig.~\ref{fig:PE}, we show the results for the BSk21 EOS, where we show the 2D and 1D
marginalized probability densities for
a subset of the recovered posterior distributions, namely, the chirp mass $\mathcal{M}$,
the mass-ratio $q$, and the dimensionless tidal deformability $\tilde{\Lambda}$. In the 2D
contour plots, we show the $90\%$ confidence intervals. The dotted lines and crosses mark 
the injected values. Through the comparison with 
\texttt{SEOBNRv5\_ROM\_NRTidalv3}\_inj\_same, we find that our employed setup is able to 
recover the injected parameters reliably and that all injected parameters are recovered 
within the 90\% confidence interval. Using our hybrid waveform all parameters are
recovered within the 90\% confidence interval, without
biases. 
We
find that the network SNR beyond which the tidal deformability deviates from its 
injected value
by more than $1 \sigma$ is $\sim $ 105. 
The lack of bias despite the tidal models not capturing the earlier onset of mass transfer
is because of the lower sensitivity of the detectors at such frequencies.
Indeed, we did a recovery with a lower upper cutoff frequency of 1792 Hz and found that
the results of the parameter estimation are almost identical, 
suggesting that the lack of post-merger content in the models
used for parameter estimation has no significant impact.
These results are in agreement with Ref.~\cite{Dudi:2018jzn}, and consistent with 
the expectation that the post-merger will have negligible impact on parameter 
recovery with the current generation of gravitational wave detectors \cite{Chatziioannou:2017ixj}.
Note that, in agreement with Refs.~\cite{Golomb:2024mmt,
LIGOScientific:2016vlm}, we find
a positive correlation between $q$ and $\tilde{\Lambda}$. Overall, our results suggest that
if we were to observe a subsolar event, current waveform models would accurately identify
the event as containing a subsolar object. 

We also show results when assuming the smaller object
is a black hole (i.e., that its tidal deformability is zero)\footnote{Note that we do not use the BHNS waveform model of the SEOBNRv4 model family, but instead force tidal deformability of the smaller object in BNS model to vanish. The reason for this is that the current BHNS model requires the larger object to be the black hole.}. We find small biases in the chirp mass and mass ratio, in addition to
the tidal deformability, although within $1 \sigma$. Finally, we also recover with
the BBH waveform model \texttt{SEOBNRv5\_ROM} and find large biases in the chirp mass and mass ratio.
The evidence for whether there is one or two neutron stars in the system is quantified with the
log Bayes factor between the BNS and BHNS model. Focusing on a SNR of 35, we find a log 
Bayes factor of 0.40 in favour of the BHNS model, meaning the BHNS model is marginally
preferred, but the evidence is inconclusive. This is consistent with Ref.~\cite{Yang:2017gfb},
which showed that the leading order tidal effects on the gravitational wave signal of an inspiraling 
compact object binary are in fact degenerate between a BNS and a BHNS binary,
so long as there is sufficient uncertainty in the EOS.
However, if we take the median recovered value for the mass of neutron star 
$M_A=1.80\ M_{\odot}$ and assume the EOS is BSk21 or SLy2, then this would imply a 
tidal deformability of 98 or 43 respectively, which is incompatible with the median
recovered value for $\Lambda_A= 990^{+380}_{-345}$. That is, no EOSs allowed by prior constraints
would give such a high value of $\Lambda$. 
Thus, once realistic EOS constraints are 
incorporated, the BHNS interpretation becomes disfavored. 
Similarly, computing the log Bayes factor between the BBH and BNS model, we find
the BNS model is strongly favoured with a log Bayes factor of 8.2.

\begin{table}[t]
\centering
\begin{tabular}{ |p{3.6cm} p{1.1cm} p{2.1cm} |   }
 \hline
        Parameter& Label & Value \\
 \hline
        Phase at 30 Hz   & $\phi$    &5.46 rad\\
        Right ascension&  $\alpha$  & 3.86 rad\\
        Declination &$\delta$ & -1.155 rad\\
        Inclination &$\iota$ & 0 rad\\
        Polarization angle &$\psi$ & 0.081 rad \\
	Merger time at geocenter &$t_c$ & 1369419318 s \\
 \hline
\end{tabular}
        \caption{Values of extrinsic parameters used for the injections.}
        \label{tab:summary_extrinsic}
\end{table}

\section{Discussion and Conclusion}
\label{sec:conclusion}
In this work, we have presented results from fully relativistic hydrodynamic simulations 
of quasi-circular BNS systems in which one of the component masses is subsolar.
We considered two different EOSs governing the perfect fluid model for the stars, 
allowing for individual tidal deformabilities up to $10^4$. 
Our focus was on quantifying the impact of 
large tidal deformabilities on the gravitational wave signal, and in this first study, we have neglected the
effects of magnetic fields, neutrinos, and realistic nuclear
microphysics.

Due to the relatively large mass ratio, we find that tidal effects are significant. 
For example, the smaller mass star can start to transfer mass to its more massive companion even during
the late inspiral.
We also find that the amount of dynamically ejected material is on 
the order of a few times $10^{-2}\ M_{\odot}$, largely independent of the EOS considered. 
This corresponds to an 
increase by a factor of $\approx 30$ compared to an equal-mass binary with the same total mass, 
in disagreement with predictions from phenomenological relations for the ejecta and disk 
masses derived from typical BNS simulations with super-solar mass neutron stars.

We next investigated the impact of large tidal deformabilities on the gravitational wave signal and found 
that matter effects lead to significantly lower disruption frequencies, on the order 
of 700 Hz---i.e., well within the sensitive band of current gravitational wave detectors. Motivated by this, 
we examined whether current gravitational wave models 
would suffer from a loss of 
sensitivity or introduce biases in the estimation of source properties. To address this, 
we constructed complete inspiral-merger-ringdown BNS waveforms by combining 
state-of-the-art EOB and IMRPhenom models for the inspiral with our numerical relativity
simulations for the late inspiral and merger. The resulting waveforms show mismatches 
with state-of-the-art models below $4 \times 10^{-3}$, indicating that such signals 
would not be missed by current searches, as the differences predominantly show up at the higher
frequencies that contribute less to the SNR.

We then injected these hybrid waveforms into fiducial data streams of the LIGO and Virgo 
detectors and performed parameter estimation using various waveform models. We found that 
neglecting all tidal effects in the waveform model leads to biases 
in the inferred chirp mass 
and mass ratio for an observation with a SNR of 35. However, the data 
would favor a BNS template over a BBH model. Conversely, assuming that the lighter object 
is a black hole introduces small biases in the recovered intrinsic parameters, and the 
data would instead favor a BHNS model. However, in that case, the inferred tidal deformability 
for the larger mass neutron star would be in tension with observations.

When recovering the signal with waveform models that include closed-form tidal corrections,
we did not find any biases in the recovered chirp mass, mass ratio, and effective tidal 
deformability, all consistent within the 90\% credible intervals. We further determined 
that the network SNR at which the inferred tidal deformability would deviate from its 
injected value by more than one standard deviation is approximately 105, corresponding 
to the upper range of expected SNRs for a GW170817-like event in the O5 observing 
run~\cite{O5}. Given the expected 
increase in detector sensitivity, our results 
suggest that current waveform systematics are unlikely to be a limiting factor for
typical O5 detections, but may become relevant for particularly loud events.
This highlights the
need for continued improvement of BNS waveform models, especially in the late inspiral, 
merger, and post-merger regimes where tidal effects become increasingly important and 
current models transition to BBH-like templates.

Overall, we hope that our results can serve as an initial testing ground for future 
gravitational-wave and electromagnetic modeling of subsolar BNS mergers. 
However, more simulations for various mass ratios, EOSs, 
and spins, as well as simulations including realistic nuclear
microphysics are needed to draw a more complete picture.

\acknowledgements
It is a pleasure to thank Samuel D. Tootle, Harald Pfeiffer, and 
Konrad Topolski for their assistance constructing 
the initial data used here; Anna Puercher, Adrian Abac, and Marcus Haberland for help
setting up the parameter estimation runs; Lami Suleiman for providing tabular equations
of state; and Alan Tsz-Lok Lam, Hao-Jui Kuan, Reed
Essick, Marcus Haberland, and Tim Dietrich for 
helpful discussions regarding various aspects of this project.
W.E. acknowledges support from a Natural Sciences and Engineering Research 
Council of Canada Discovery Grant and
an Ontario Ministry of Colleges and Universities Early Researcher Award.
This research was
supported in part by Perimeter Institute for Theoretical Physics. Research at
Perimeter Institute is supported in part by the Government of Canada through
the Department of Innovation, Science and Economic Development and by the
Province of Ontario through the Ministry of Colleges and Universities. 
Computations were performed on the
Urania and Hypatia HPC system at the Max Planck Computing and Data Facility.
This material is based upon work supported by NSF's LIGO Laboratory which is a major 
facility fully funded by the National Science Foundation.
\newpage
\appendix
\section{\label{app:convergence} Numerical convergence and error estimates}

\begin{figure*}
        \includegraphics[width=0.99\columnwidth,draft=false]{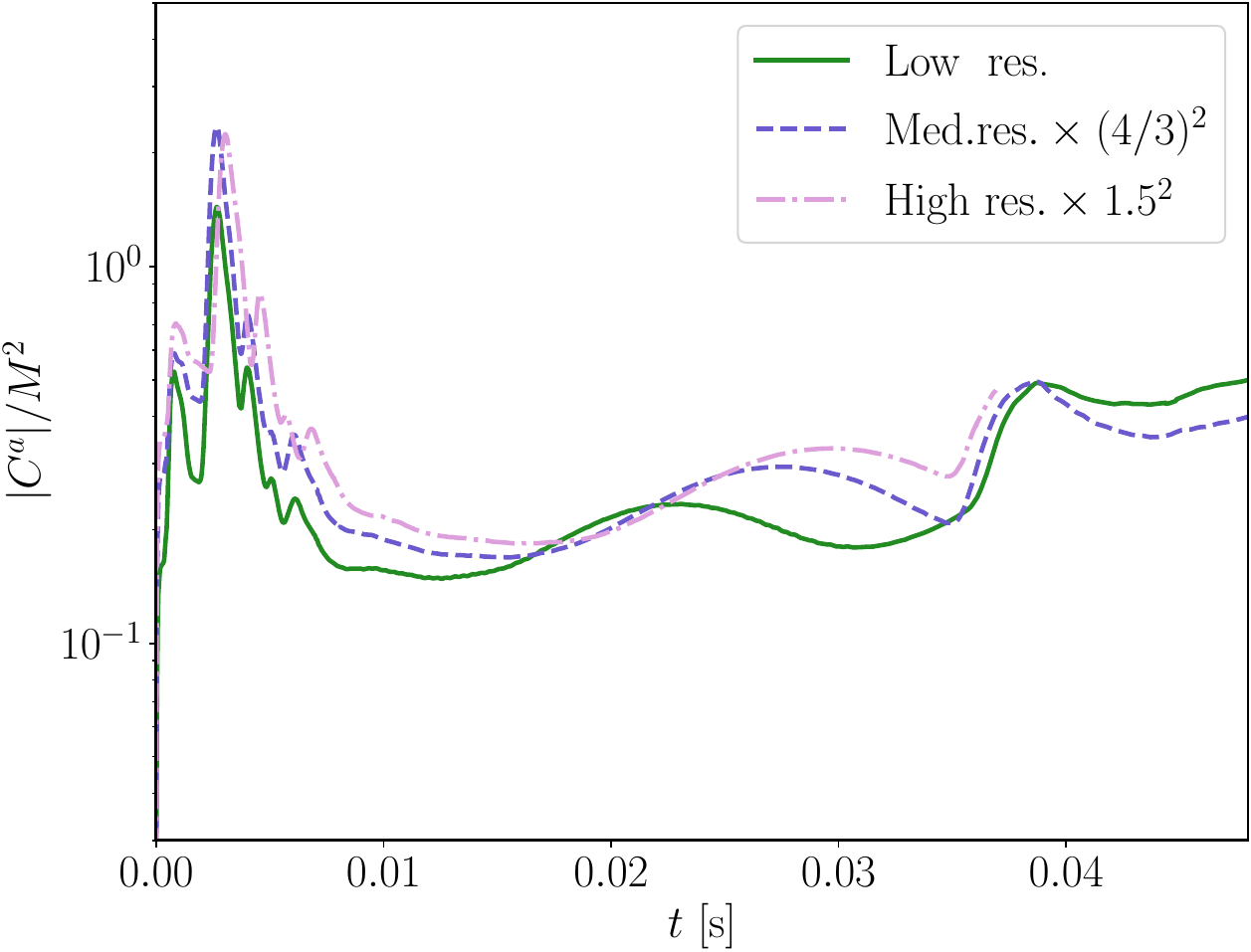}
        \includegraphics[width=0.99\columnwidth,draft=false]{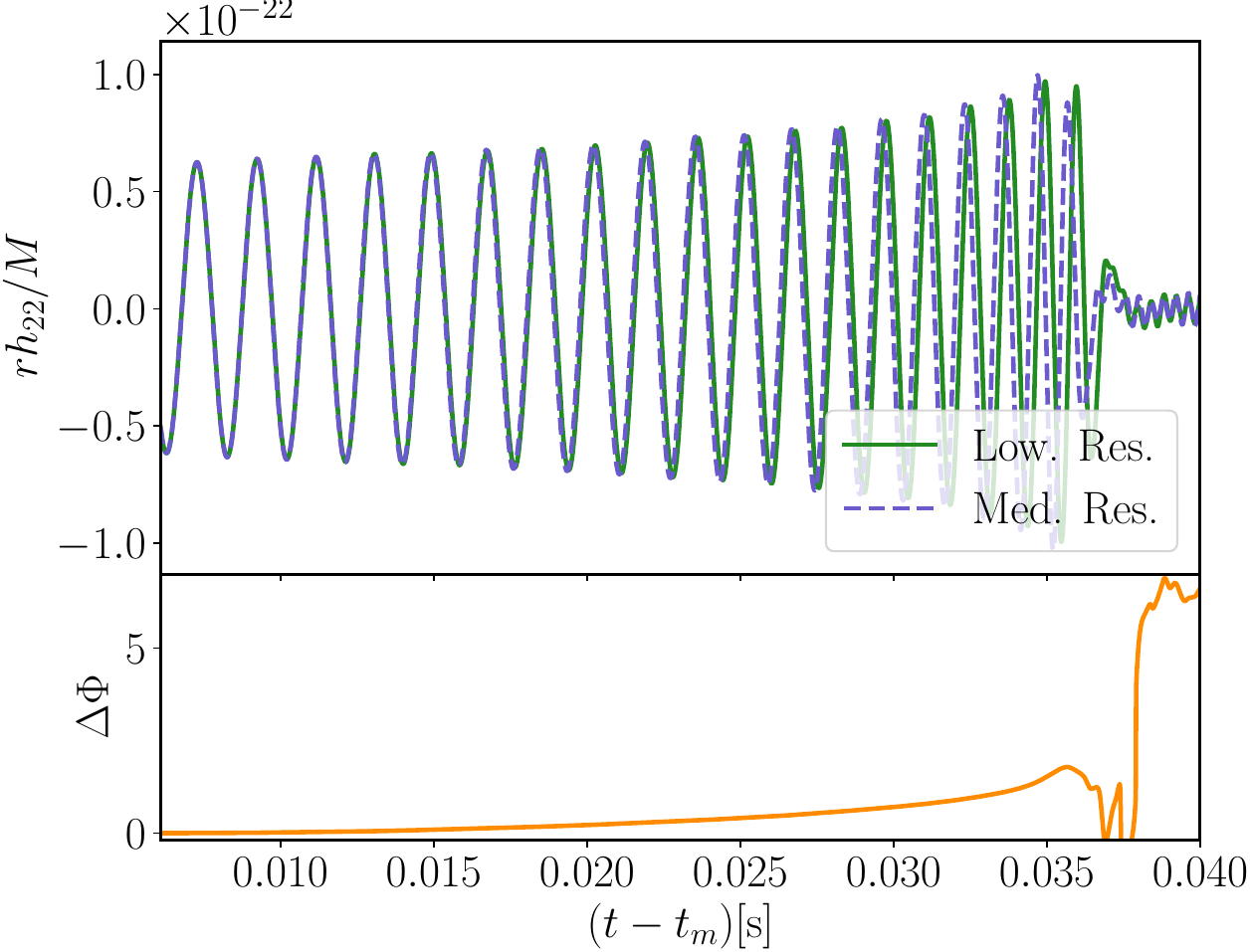}
        \caption{Left: Volume integrated norm of the
        generalized harmonic constraint violation $C^a$ as a function
        of time for the BNS system for BSk21 EOS. The values
        have been scaled assuming second order convergence, though at early times the
        convergence is closer to first order. Right: Gravitational wave radiation from the low 
	and medium resolutions aligned
	in phase and time at $500$ Hz (top) and the corresponding accumulated dephasing 
	between the low and medium resolutions (bottom).
\label{fig:cnst}
}
\end{figure*}

For the BNS mergers considered in this paper, we perform
simulations with seven levels of refinement where the finest level has a linear grid 
spacing of $dx \sim 0.033 M$,
and each successive level has a linear grid spacing that is twice as coarse.
In the left panel of Fig.~\ref{fig:cnst}, we show the norm of the generalized harmonic
constraint violation, integrated over the domain as a function of time.
For the BSk21 EOS, we also
perform a convergence study with grid spacing that is $3/4$ and $2/3 \times$ as large,
which is shown in Fig.~\ref{fig:cnst}.
All results in the main text are from the lowest resolution.
Although at early times, the order of convergence is closer to first order,
presumably from high frequency noise (junk radiation) in the initial data
which may engage the shock capturing scheme,
at later times, the convergence is consistent
with roughly second order, as expected from our numerical scheme in the absence of shocks.
The right panel of Fig.~\ref{fig:cnst} shows the waveforms for the low and medium resolutions (top panel)
and the dephasing (bottom panel) which accumulates to roughly 1.78 radians at merger,
defined as the time at which the complex amplitude of the waveform peaks.

\begin{figure*}
        \includegraphics[width=0.99\columnwidth,draft=false]{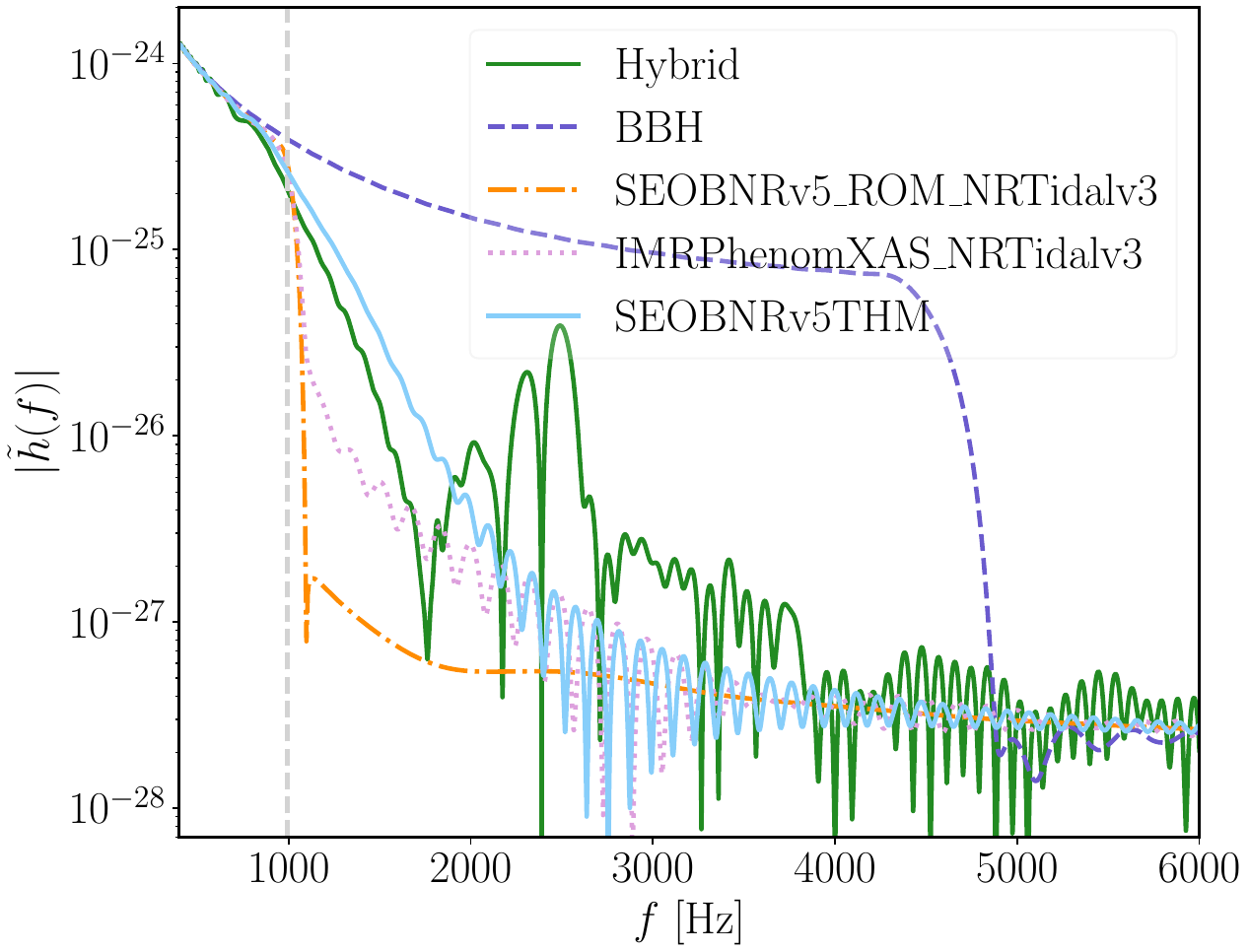}
        \includegraphics[width=0.99\columnwidth,draft=false]{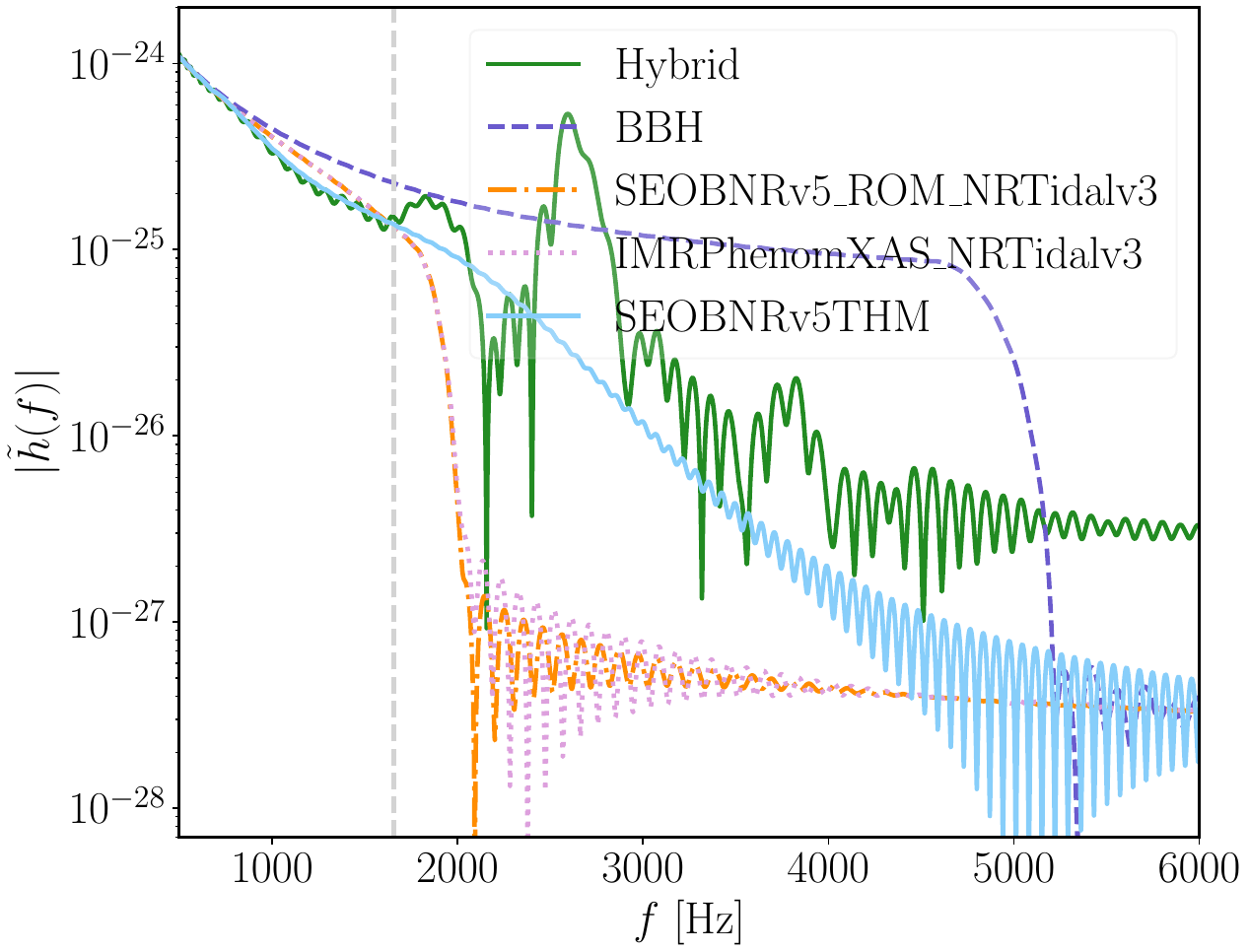}

        \caption{The frequency domain hybridized and analytical waveforms as a function
        of frequency for the BSk21 EOS. The left panel shows the unequal mass system
        presented in main text. The right panel shows the results when evolving an equal
        mass system with the same total mass. 
        The dashed, gray lines mark the merger frequency.
        We find that the agreement is better up to larger
        frequencies for the equal mass case, because the tidal effects are smaller.
\label{fig:htilde}
}
\end{figure*}

\begin{figure*}
        \includegraphics[width=0.99\columnwidth,draft=false]{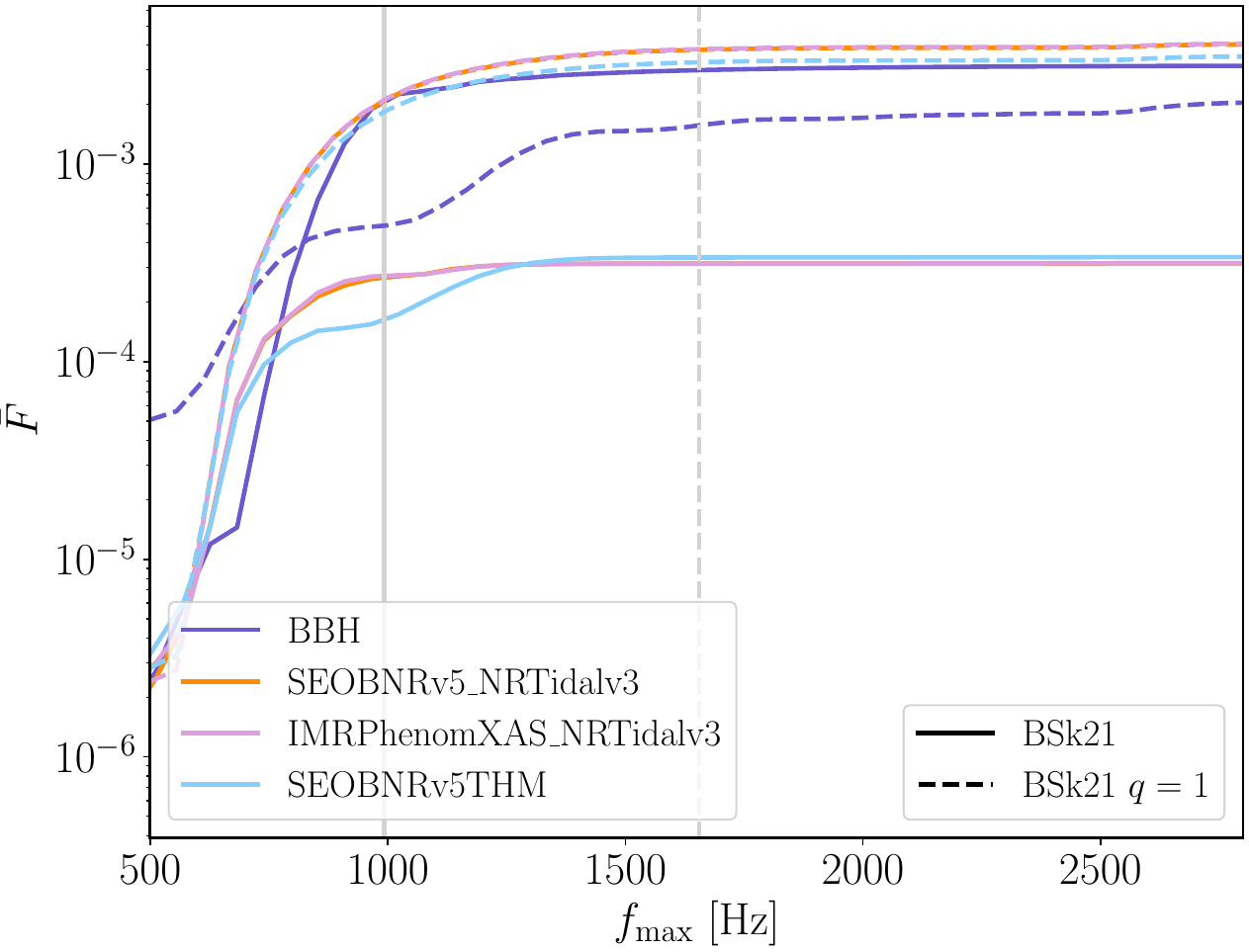}
        \includegraphics[width=0.99\columnwidth,draft=false]{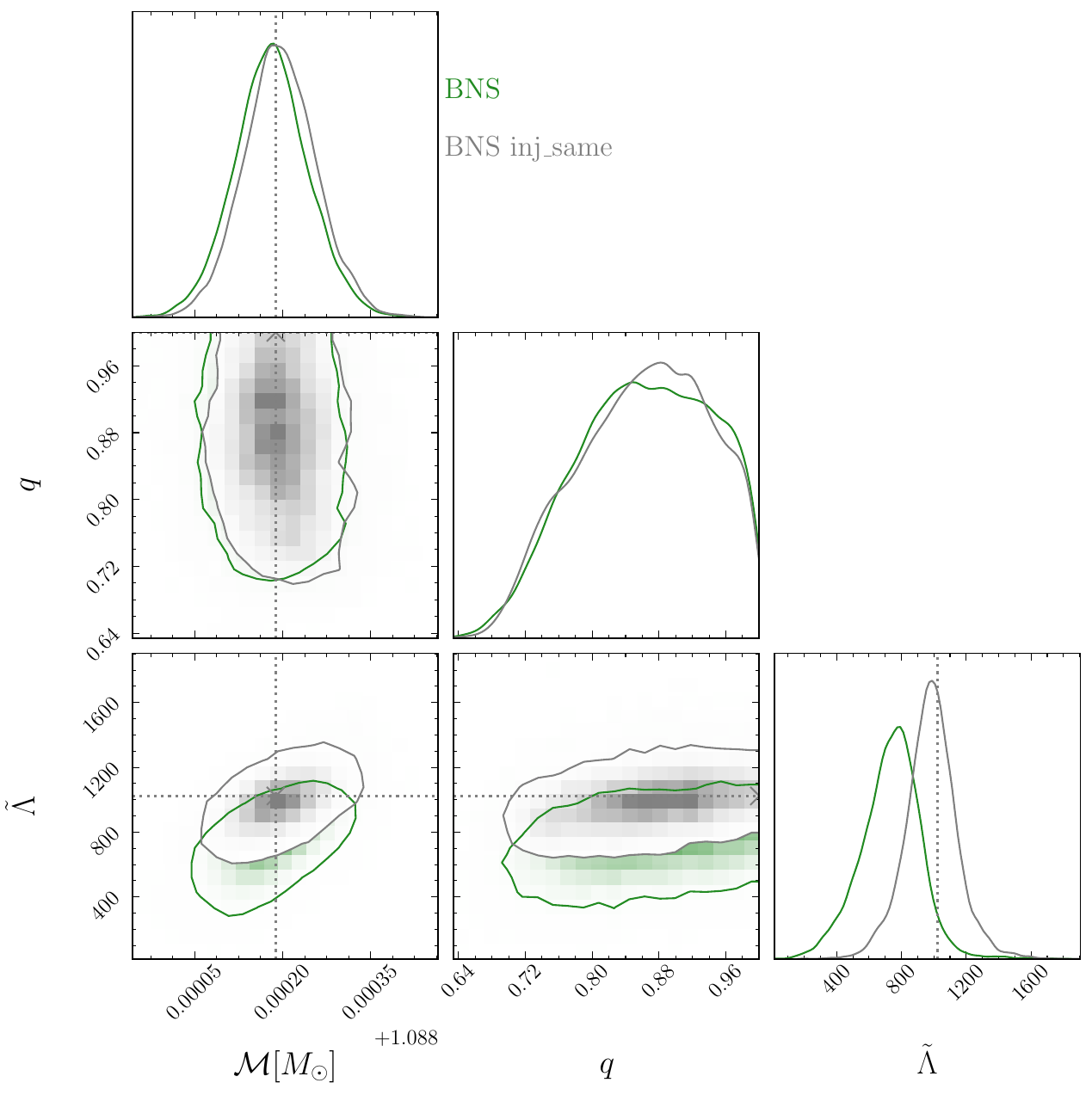}

	\caption{Left: Mismatches between different waveform models and our hybrid 
	waveforms computed as in Fig.~\ref{fig:mismatches}, but for
	the equal mass (dashed lines) and unequal mass (solid lines) with the BSk21 EOS. Right: 
	The marginalized 1D and 2D posterior probability distributions for chirp mass $\mathcal{M}$, mass ratio $q$, and tidal deformability $\tilde{\Lambda}$ for the
        equal mass BSk21 configuration. The recovery labeled by BNS (green) uses
a different approximation to recover our hybridized waveform which is a combination of the
\texttt{SEOBNRv5\_ROM\_NRTidalv3} model and our numerical simulation. The
BNS inj\_same (gray) uses for injection and recovery the \texttt{SEOBNRv5\_ROM\_NRTidalv3} model with the same source parameters as the hybrid.
\label{fig:mismatch_q_dependence}
}
\end{figure*}

As a consistency check, we perform simulations for an equal mass system with the same
total mass. Again, focusing on the BSk21 EOS, we hybridize the waveform as described
in the main text. In Fig.~\ref{fig:htilde}, we present the frequency domain hybridized
and analytical waveforms. We find that the equal mass system agrees with analytical
waveforms up to larger frequencies. This is confirmed by computing the mismatches
between hybrid and analytical waveforms, which we show for both the equal and unequal
mass case in Fig.~\ref{fig:mismatch_q_dependence}.
From this figure, one can also see that the gravitational wave signal in the 2--3 kHz frequency range,
coming from the post-merger oscillations, is significantly weaker in the unequal
mass case compared to the equal mass case. 

As a final consistency check, we perform parameter estimation on the hybrid waveform for 
the equal mass case. We choose a distance that gives a similar SNR to the one considered 
in the main text. We use similar priors and show the results in Fig.~\ref{fig:mismatch_q_dependence}.
As mentioned above, the masses and spins are well-recovered; however, although the injected
value of the effective tidal deformability is within the 90\% confidence interval, 
there is a small bias
towards lower effective tidal deformability. This is consistent 
with the results found in Ref.~\cite{Dudi:2018jzn}, and is likely due to the accumulated phase 
difference between our hybrid and analytical waveform in the late inspiral. This is
further justified by the observation that the posteriors are nearly identical when
analyzing the signal with a lower cutoff frequency, again in agreement with 
with Ref.~\cite{Dudi:2018jzn}.

\section{\label{app:mismatches} Normalization Effects and Alternative Overlap Measures}
In this Appendix, we present additional diagnostics to clarify 
why, for the BSK21 EOS, the BBH approximant \texttt{SEOBNRv5\_ROM} yields smaller mismatches 
than the tidal models before the onset of significant mass transfer.

\begin{figure*}
        \includegraphics[width=0.99\columnwidth,draft=false]{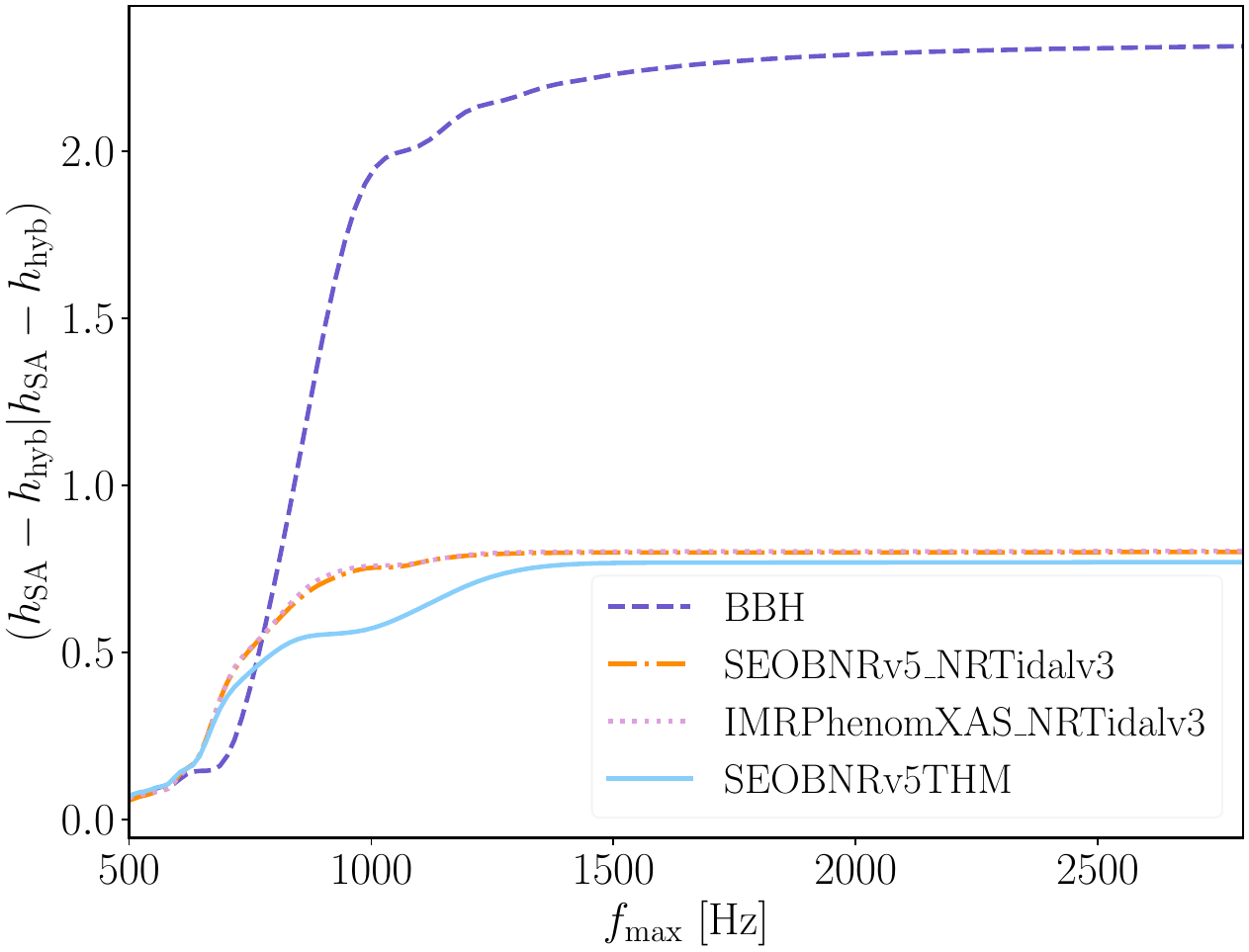}
        \includegraphics[width=0.99\columnwidth,draft=false]{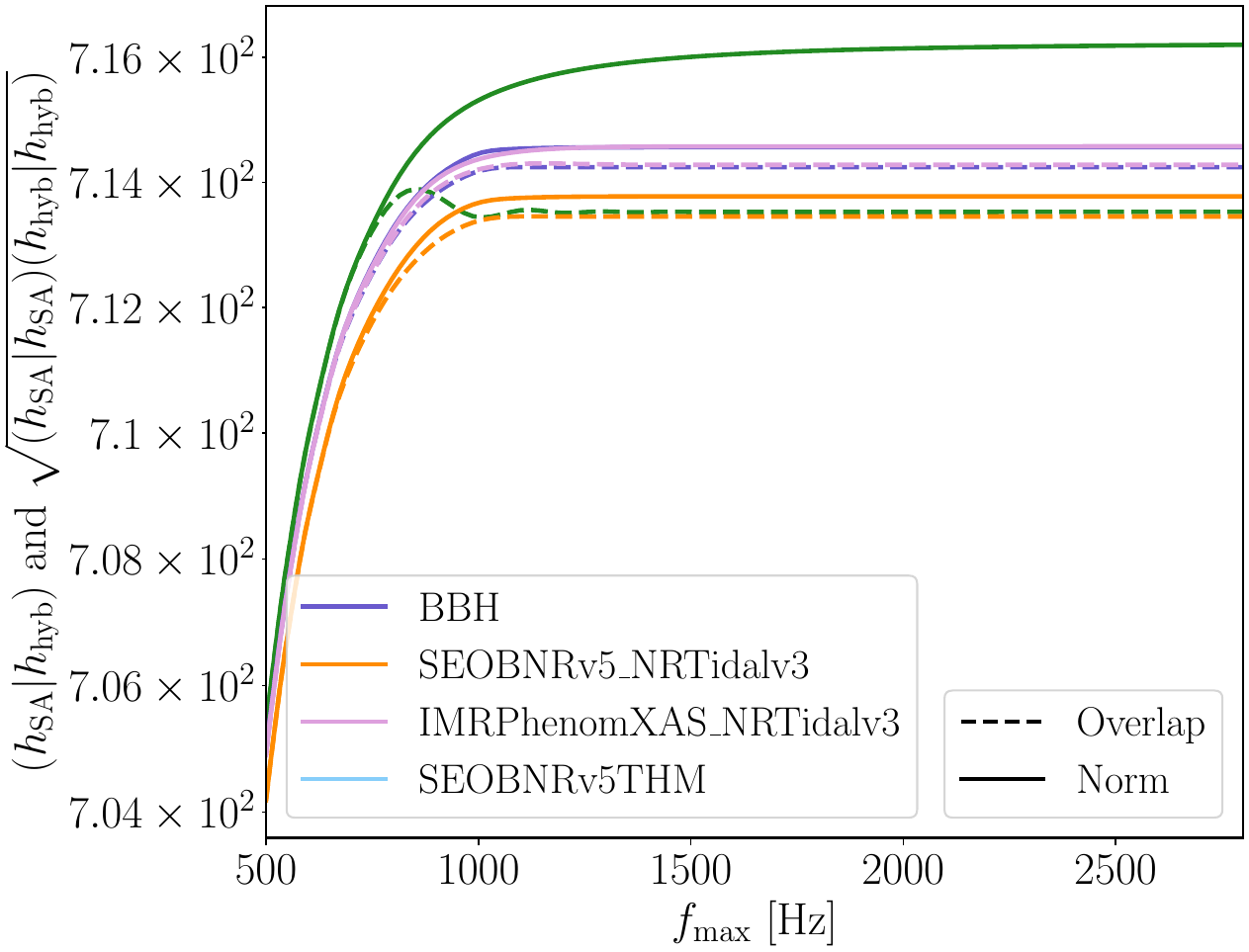}

	\caption{Left: A match-based estimate of the size of waveform differences at a 
	fixed distance of $100$Mpc, as a function
	of upper cutoff frequency for the BSK21 EOS. Right: The un-normalized overlap between the 
	semi-analytical and hybrid waveforms $(h_{\rm SA}|h_{\rm hyb})$, and the 
	geometric mean amplitude of the 
	two signals $(h_{\rm SA/hyb}|h_{\rm SA/hyb})$, 
	as a function of the upper cutoff frequency, for the BSK21 EOS.
\label{fig:waveform_diff}
}
\end{figure*}

While the standard mismatch 
provides a convenient and widely used measure of waveform agreement, it involves an 
implicit normalization that can obscure the relative impact of amplitude differences, 
particularly in regimes where tidal effects are still subdominant. To better disentangle 
normalization effects from genuine waveform disagreement, we therefore consider 
alternative measures that more directly quantify the noise-weighted difference between 
waveforms. 
First, in the left panel of Fig.~\ref{fig:waveform_diff} we show the quantity
\begin{equation}
(h_{\rm SA}-h_{\rm hyb}|h_{\rm SA}-h_{\rm hyb}),
\end{equation}
evaluated at a fixed reference distance of $100$ Mpc. 
This measure more directly maps onto the likelihood weighting in parameter estimation, 
since it quantifies the noise-weighted squared difference between signals without the 
additional normalization inherent in the mismatch. 
We find that in the low-frequency regime, where the BBH approximant also exhibits slightly 
smaller mismatches, the squared difference is correspondingly small, indicating that 
tidal contributions to the signal remain subdominant there. At higher frequencies, 
however, the tidal models show improved agreement as expected once tidal effects become 
significant.

Second, in the right panel of Fig.~\ref{fig:waveform_diff}, we show the 
un-normalized overlap between the semi-analytical and hybrid waveforms,
and the geometric mean amplitude of the two signals, as a function of the upper 
cutoff frequency. This figure makes explicit how the cumulative disagreement 
builds up with frequency while avoiding the rescaling present in the standard mismatch 
definition. This again indicates that the apparent better agreement at low-frequencies of
the BBH model is primarily driven by normalization effects, rather than improved 
physical agreement with the hybrid waveform.

\bibliography{./main}

\end{document}